\documentclass[twocolumn]{aastex61}
\pdfoutput=1 
\usepackage{amsmath,amstext,amssymb}
\usepackage[T1]{fontenc}
\usepackage{apjfonts} 
\usepackage[figure,figure*]{hypcap}
\usepackage{color}
\usepackage{longtable}

\shorttitle{WASP-52$\MakeLowercase{{\rm b}}$}
\shortauthors{Alam et al.}

\begin{document}

\title{The \textit{HST} P\MakeLowercase{an}CET Program: Hints of N\MakeLowercase{a} {\sc i} \& Evidence of a Cloudy Atmosphere for the Inflated Hot Jupiter WASP-52$\MakeLowercase{{\rm b}}$}

 \author{Munazza K. Alam}
 \altaffiliation{National Science Foundation Graduate Research Fellow}
\affiliation{Department of Astronomy, Harvard-Smithsonian Center for Astrophysics, Cambridge, MA 02138, USA}
\correspondingauthor{Munazza Alam}
\email{munazza.alam@cfa.harvard.edu}

\author{Nikolay Nikolov}
\affiliation{Astrophysics Group, School of Physics and Astronomy, University of Exeter, Exeter EX4 4QL, UK}

\author{Mercedes L\'opez-Morales}
\affiliation{Department of Astronomy, Harvard-Smithsonian Center for Astrophysics, Cambridge, MA 02138, USA}

\author{David K. Sing}
\affiliation{Astrophysics Group, School of Physics and Astronomy, University of Exeter, Exeter EX4 4QL, UK}
\affiliation{Department of Earth and Planetary Sciences, Johns Hopkins University, Baltimore, MD 21218, USA}

\author{Jayesh M. Goyal}
\affiliation{Astrophysics Group, School of Physics and Astronomy, University of Exeter, Exeter EX4 4QL, UK}

\author{Gregory W. Henry}
\affiliation{Center of Excellence in Information Systems, Tennessee State University, Nashville, TN  37209, USA}

\author{Jorge Sanz-Forcada}
\affiliation{Centro de Astrobiolog\'ia (CSIC-INTA), ESAC Campus, Villanueva de la Ca\~nada, Madrid, Spain}

\author{Michael H. Williamson}
\affiliation{Center of Excellence in Information Systems, Tennessee State University, Nashville, TN  37209, USA}

\author{Thomas M. Evans}
\affiliation{Astrophysics Group, School of Physics and Astronomy, University of Exeter, Exeter EX4 4QL, UK}

\author{Hannah R. Wakeford}
\affiliation{Space Telescope Science Institute, Baltimore, MD 21218, USA}

\author{Giovanni Bruno}
\affiliation{Space Telescope Science Institute, Baltimore, MD 21218, USA}
\affiliation{INAF-Osservatorio Astrofisico di Catania, via S. Sofia, 78, 95123 Catania, Italy}

\author{Gilda E. Ballester}
\affiliation{Lunar and Planetary Laboratory, University of Arizona, Tucson, AZ 85721, USA}

\author{Kevin B. Stevenson}
\affiliation{Space Telescope Science Institute, Baltimore, MD 21218, USA}

\author{Nikole K. Lewis}
\affiliation{Space Telescope Science Institute, Baltimore, MD 21218, USA}
\affiliation{Department of Astronomy and Carl Sagan Institute, Cornell University, Ithaca, NY 14853, USA}

\author{Joanna K. Barstow}
\affiliation{Department of Physics and Astronomy, University College London, London WC1E 6BT, UK}

\author{Vincent Bourrier}
\affiliation{Observatoire de l'Universit\'e de Gen\`eve, Sauverny, Switzerland}

\author{Lars A. Buchhave}
\affiliation{DTU Space, National Space Institute, Technical University of Denmark, Lyngby, Denmark}

\author{David Ehrenreich}
\affiliation{Observatoire de l'Universit\'e de Gen\`eve, Sauverny, Switzerland}

\author{Antonio Garc\'ia Mu\~noz}
\affiliation{Zentrum f\"ur Astronomie und Astrophysik, Technische Universit\"at Berlin, Berlin, Germany}

\begin{abstract}
We present an optical to near-infrared transmission spectrum of the inflated hot Jupiter WASP-52b using three transit observations from the Space Telescope Imaging Spectrograph (STIS) mounted on the \textit{Hubble Space Telescope}, combined with \textit{Spitzer}/Infrared Array Camera (IRAC) photometry at 3.6 $\mu$m and 4.5 $\mu$m. Since WASP-52 is a moderately active (log($L_{\rm x}/L_{\rm bol}$) = $-$4.7) star, we correct the transit light curves for the effect of stellar activity using ground-based photometric monitoring data from the All-Sky Automated Survey for Supernovae (ASAS-SN) and Tennessee State University's Automatic Imaging Telescope (AIT). We bin the data in  38 spectrophotometric light curves from 0.29 to 4.5 $\mu$m and measure the transit depths to a median precision of 90 ppm. We compare the transmission spectrum to a grid of forward atmospheric models and find that our results are consistent with a cloudy spectrum and evidence of sodium at 2.3$\sigma$ confidence, but no observable evidence of potassium absorption even in the narrowest spectroscopic channel. We find that the optical transmission spectrum of WASP-52b is similar to that of the well-studied inflated hot Jupiter HAT-P-1b, which has comparable surface gravity, equilibrium temperature, mass, radius, and stellar irradiation levels. At longer wavelengths, however, the best fitting models for WASP-52b and HAT-P-1b predict quite dissimilar properties, which could be confirmed with observations at wavelengths longer than $\sim$1 $\mu$m. The identification of planets with common atmospheric properties and similar system parameters will be insightful for comparative atmospheric studies with the \textit{James Webb Space Telescope}.

\end{abstract}


\section{Introduction}
\label{sec:intro}

Transiting exoplanets offer unprecedented opportunities for the detection and detailed characterization of planets beyond the Solar System \citep{Struve52}. From transit observations, we can make inferences about the formation and evolutionary histories of these planets, their bulk compositions, and their atmospheres \citep{Winn10}. Atmospheric studies can be performed using transmission spectroscopy to constrain atmospheric structure and chemical composition \citep[e.g.,][]{Charbonneau02,Vidal-Madjar03,Wakeford17}, secondary eclipses to measure temperature and thermal structure \citep[e.g.,][]{Deming05,Charbonneau08,SingLM09,Evans17}, and orbital phase curves to probe atmospheric circulation \citep[e.g.,][]{Knutson07a,Stevenson17}. The subject of this paper is transmission spectroscopy \citep{Seager00,Brown01}. During transit, light from the host star passes through the atmosphere of the planet. At wavelengths where absorption by atoms, molecules, and aerosols takes place, the planet blocks slightly more stellar flux, resulting in variations in the apparent radius of the planet as a function of wavelength. These variations in planetary radius reveal the composition of the planetary atmosphere.

The gaseous atmospheres of short-period hot Jupiters are most accessible to such observations because of their large scale heights and short orbital periods. Narrow peaks of H {\sc i} (1215 \AA) in the UV, Na {\sc i} (5893 \AA) and K {\sc i} (7665 \AA) in the optical, H$_{2}$O (1.4 $\mu$m; 1.9 $\mu$m), CO (2.3 $\mu$m; 4.7 $\mu$m), CO$_{2}$ (2 $\mu$m; 15 $\mu$m), and CH$_{4}$ (2.2 $\mu$m; 7.5 $\mu$m) in the near-infrared, and scattering by molecular hydrogen (H$_{2}$) are expected to be prominent features in clear hot Jupiter atmospheres \citep{Vidal-Madjar03,Seager00,Sudarsky03,Burrows10,Fortney10}. \citet{Charbonneau02} detected the first exoplanet atmosphere for the hot Jupiter HD 209458b, which was later confirmed to have absorption from Na {\sc i}, H$_{2}$ Rayleigh scattering, and possible TiO/VO absorption \citep[][]{Sing08a,Snellen08,Lecavelier08,Desert08,Sing08b}. The first near-UV to IR (0.3-8.0 $\mu$m) transmission spectrum of the hot Jupiter HD 189733b \citep{Pont13} revealed clouds/hazes consistent with Rayleigh scattering by small condensate particles in addition to narrow peaks of Na {\sc i} and K {\sc i}, H$_{2}$O absorption, and an escaping H atmosphere \citep[]{Grillmair08,Sing09,Lecavelier10,Bourrier13}. 

To date, a diversity of hot Jupiters with a continuum of clear to cloudy atmospheres \citep{Sing16} has been detected, with no apparent correlation between the observed spectra and other system parameters. Space-based atmospheric studies have yielded detections of Na {\sc i} and K {\sc i} (e.g., \citealt{[]Charbonneau02,Nikolov14}), elucidated the presence of thick atmospheric cloud decks (e.g., \citealt{Sing15}), and have provided water abundance constraints (e.g., \citealt{[]Kreidberg14,Wakeford18}). 
From ground-based transmission spectral surveys of hot Jupiters, we have detected Na {\sc i} \citep[e.g.,][]{Nikolov16,Nikolov18,Wyttenbach15,Wyttenbach17}, K {\sc i} \citep[e.g.,][]{Sing11a,Nikolov16}, cloudy/hazy atmospheres \citep[e.g.,][]{Jordan13,Mallonn16,Huitson17}, and Rayleigh scattering slopes \citep[e.g.,][]{Gibson17}.


\capstartfalse

\begin{deluxetable*}{cccccc}
\tabletypesize{\scriptsize}
\tablewidth{0pt}
\tablecolumns{7}
\tablecaption{Transit Observations of WASP-52b\label{tab:obs}}

\tablehead{\colhead{Obs Date} & \colhead{Visit Number} & \colhead{Telescope/Instrument} & \colhead{Grating/Grism}  & \colhead{Number of Images} & \colhead{Exposure Time (sec)}} 

\startdata 
UT 2016 Nov 01  &  52 & \textit{HST}/STIS  &  G430L\tablenotemark{a}  &   37  &  253  \\
UT 2016 Nov 29  &  53 & \textit{HST}/STIS  &  G430L\tablenotemark{a}  &   37  &  253  \\
UT 2017 May 11  &  54 & \textit{HST}/STIS  &  G750L\tablenotemark{b}  &   37  &  253  \\
UT 2016 Oct 18  & $-$  & \textit{Spitzer}/IRAC  &  [3.6 $\mu$m]	       &   28800   &  1.92     \\
UT 2018 Mar 22  & $-$  & \textit{Spitzer}/IRAC  &  [4.5 $\mu$m]	       &   29300   &   1.98    \\
\enddata
\tablenotetext{a}{Central Wavelength: 4300 \AA}
\tablenotetext{b}{Central Wavelength: 7751 \AA}

\end{deluxetable*}
\label{tab:obs}
\capstarttrue

Here we present results for WASP-52b \citep{Hebrard13} from the \textit{Hubble Space Telescope} (\textit{HST}) Panchromatic Comparative Exoplanetology Treasury (PanCET) program (GO 14767; PIs Sing \& L\'{o}pez-Morales). The scientific goals of PanCET are to provide a uniform, statistically compelling ultraviolet (UV) through infrared (IR) study of clouds/hazes and chemical composition in exoplanet atmospheres, and assemble a UVOIR legacy sample of exoplanet transmission spectra that will be well-suited for follow-up with the \textit{James Webb Space Telescope (JWST)}.

WASP-52b is a 0.46 $M_{\rm Jup}$ and 1.27 $R_{\rm Jup}$ inflated hot Jupiter ($T_{\rm eq}$ = 1300 K) orbiting a moderately active ($\rm log(R'_{HK}) = -4.4 \pm 0.2$, \citealt{Hebrard13}; $\rm log(L_{\rm x}/L_{\rm bol})$ = $-$4.7, Section \ref{sec:activity_levels}) K2V star with a period of 1.75 days \citep{Hebrard13}. This planet, at a spectroscopic parallax distance of 175.7$\pm$1.3 pc \citep{Gaia18}, is a favorable target for atmospheric studies via transmission spectroscopy due to its large scale height ($H$ = 700 km) and deep transit ($\delta$ = 0.028). Based on its surface gravity (log($g$) = 2.87 dex) and equilibrium temperature ($T_{\rm eq}$ = 1315 K), WASP-52b is predicted to have a predominantly cloudy atmosphere with muted spectral features \citep{Stevenson16}. 
However, two recent ground-based atmospheric analyses of this target claim discrepant conclusions. \citet{Louden17} cite an optically thick cloud deck to explain an observed flat transmission spectrum, but note that their results are inconsistent with deeper transit depths at longer wavelengths \citep{Kirk16}. Conversely, \citet{Chen17} report a cloudy atmosphere with a noticeable Na {\sc i} detection and a weaker detection of K {\sc i} absorption. 

In this work, we measure WASP-52b's transmission spectrum over the $\sim$0.29$-$4.5 $\mu$m wavelength range by combining \textit{HST}/STIS and \textit{Spitzer}/IRAC observations. An outline of the paper is as follows. In Section \ref{sec:obs}, we describe the observations and data reduction techniques. In Section \ref{sec:activity_corr}, we detail the stellar activity correction. The light curve fits and measurement of the transmission spectrum are described in Section \ref{sec:analysis}. We compare our results to previously published measurements and to a grid of forward atmospheric models in Section \ref{sec:results}, and present an interpretation of the transmission spectrum in Section \ref{sec:discussion}. We summarize the paper in Section \ref{sec:conclusions}. 

\section{Observations \& Data Reduction}
\label{sec:obs}

\subsection{Observations}

We obtained time series spectroscopy during three transits of WASP-52b with \textit{HST}/STIS on UT 2016 November 01 and UT 2016 November 29 using the G430L grating (2892-5700 \AA) and UT 2017 May 11 using the G750L grating (5240-10270 \AA). 
Two additional transits were observed with \textit{Spitzer}/IRAC as part of GO program 13038 (PI Stevenson) in the 3.6 $\mu$m channel on UT 2016 October 18 and in the 4.5  $\mu$m channel on UT 2018 March 22. 
Table~\ref{tab:obs} summarizes the transit observations and instrument settings for each visit. 

\subsubsection{HST/STIS}
The G430L and G750L STIS data have resolving power R$\sim$500 and each consist of 37 stellar spectra taken over four, consecutive 96-minute orbits.  The visits were scheduled to include the transit event in the third orbit, providing an out-of-transit baseline time series before and after the transit as well as good coverage between second and third contact. We used a 128 pixel wide subarray and exposure times of 253 seconds to reduce the readout times between exposures. To minimize slit losses, the data were taken with the 52 x 2 arcsec$^{2}$ slit. 

\subsubsection{Spitzer/IRAC}
Two transits of WASP-52b were observed on UT 2016 October 18 and UT 2018 March 22 with the {\it{Spitzer}} space telescope \citep{Werner04} using the $3.6~\mu$m and $4.5~\mu$m IRAC channels \citep{Fazio04}. Although these observations cover the complete phase curve of the planet, for this work we only use a 6-hour portion of the phase curve centered on the transit event with enough out-of-transit baseline flux to allow for accurate analysis. Each IRAC exposure had an effective integration time of $\sim$2 seconds, resulting in $\sim$30,000 images for the portion of the phase curve corresponding to the transit. 


\subsection{Data Reduction}

\subsubsection{HST/STIS}
We reduced (bias-, dark- and flat-corrected) the raw 2D G430L and G750L spectra using the {\tt{CALSTIS}}\footnote{\url{http://www.stsci.edu/hst/stis/software/analyzing/calibration/pipe_soft_hist/intro.html}} pipeline (version 3.4) and the relevant calibration frames. Following the procedure detailed in \citet{Nikolov14}, we used median-combined difference images to identify and correct for cosmic ray events and bad pixels flagged by {\tt{CALSTIS}}. We extracted 1D spectra from the calibrated {\tt{.flt}} science files using IRAF's {\tt{APALL}} task. To identify the most appropriate aperture, we extracted light curves of aperture widths ranging from 6 to 18 pixels with a step size of 1. We defined the best aperture for each grating according to the lowest photometric dispersion in the out-of-transit baseline flux. Based on this criterion, we used an aperture size of 13 pixels in our analysis. For each exposure, we computed the mid-exposure time in MJD.

Aperture extractions with no background subtraction minimize the out-of-transit standard deviation of the white light curves \citep{Sing15}. Although the STIS 2D spectra are known to show negligible background sky contribution (\citealt[]{Sing11, Sing13, Huitson12, Huitson13, Nikolov15, Gibson17}), we assessed the potential bias of the sky background on the light curves by obtaining time series spectroscopy both with and without background subtraction.  Comparing the light curves, we find that both data sets are fully consistent. To obtain a wavelength solution, we used the {\tt{x1d}} files from {\tt{CALSTIS}} to re-sample all of the extracted spectra and cross-correlate them to a common rest frame. The cross-correlation measures the shift, and the spectra are re-sampled to align them and remove sub-pixel drifts in the dispersion direction. These drifts can be associated with the different locations of the spacecraft on its orbit around the Earth (e.g., \citealt{Huitson13}).  

\subsubsection{Spitzer/IRAC}
We analyzed the IRAC photometry following the methodology of \citet{Nikolov15} and \citet{Sing15,Sing16}. We started our analysis using the Basic Calibrated Data ({\tt{.bcd}}) files and converted the images from flux in mega-Jansky per steradian (MJy sr$^{-1}$) to photon counts (i.e., electrons) by multiplying each image by the gain and exposure time and then dividing by the flux conversion factor. Following the reduction procedure outlined in \citet{Knutson12} and \citet{Todorov13}, we filtered for outliers (hot or lower pixels in the data) by following each pixel in time. We scanned the images in two passes: first by removing outliers $\geq$$8\sigma$ away from the median value of each frame compared to the 10 surrounding images and then by removing outliers above the $4\sigma$ level following the same procedure. The total fraction of corrected pixels was $0.05\%$. 

We estimated and subtracted the sky background for each image using an iterative $3\sigma$ clipping procedure in which we excluded all pixels associated with the stellar point spread function (PSF), background stars, or hot pixels. In the last iteration, we created a histogram from the remaining pixels and determined the sky background based on a Gaussian fit to the distribution of remaining pixels. To locate the center of the PSF for each image, we used the flux-weighted centroiding method with a 5-pixel radius circular region centered on the approximate position of the star. The variation of the $x$ and $y$ positions of the PSF on the detectors were measured to be 0.19 and 0.24 pixels, respectively.

We extracted photometric points following fixed and time-variable photometry. In the fixed approach, we used circular apertures ranging in radius from 4 to 8 pixels in increments of 0.5. For the time-variable photometry, the aperture size was scaled by the value of the noise pixel parameter (the normalized effective background area of the IRAC point response function), which depends on the full-width-at-half-maximum (FWHM) of the stellar PSF squared \citep[]{Mighell05,Knutson12,Lewis13,Nikolov15}. We identified the best results from both photometric methods by comparing the light curve residual dispersion, as well as the white and red noise components measured with the wavelet technique detailed in \citet{Carter09}. The time-variable method resulted in the lowest white and random red noise correlated with data points co-added in time.

\capstartfalse

\begin{deluxetable}{ccccc}
\tabletypesize{\scriptsize}
\tablewidth{0pt}
\tablecolumns{5}
\tablecaption{Summary of Photometric Observations for WASP-52b\label{tab:activity}}

\tablehead{\colhead{Observing} & \colhead{$N_{obs}$} & \colhead{Date Range} & \colhead{Sigma}  & \colhead{Seasonal Mean} \\ \colhead{Season} & \colhead{} & \colhead{(HJD - 2,450,000)} & \colhead{(mag)} & \colhead{(mag)} } 

\startdata 
\cutinhead{AIT}
2014$-$2015 & 61  & 56943$-$57066  & 0.14288   & 0.01270  \\
2015$-$2016 & 48  & 57293$-$57418  & 0.15069   & 0.00892  \\
2016$-$2017 & 27  & 57705$-$57785  & 0.16062   & 0.01335  \\
\cutinhead{ASAS-SN} 
2013$-$2014 & 154  & 56638$-$57031  & 0.16692 & 0.01162 \\
2014$-$2015 & 175  & 57145$-$57385  & 0.15601 & 0.01558 \\
2015$-$2016 & 209  & 57507$-$57749  & 0.11866 & 0.01195 \\
2016$-$2017 & 202  & 57878$-$58115  & 0.11769 & 0.01167\\
\enddata

\end{deluxetable}
\label{tab:activity}
\capstarttrue

\section{Stellar Activity Correction}	
\label{sec:activity_corr}

Since WASP-52 is a moderately active star ($\rm log(R'_{HK}) = -4.4 \pm 0.2$, \citealt{Hebrard13}), we used ground-based activity monitoring data to track stellar activity levels during the epochs of our transits. Before fitting the transit light curves, we corrected the baseline flux levels for the effect of stellar variability using a quasi-periodic Gaussian process regression model and corrected for the effect of unocculted stellar spots following the prescription of \citet{Huitson13}. Table~\ref{tab:activity} summarizes the photometric monitoring campaigns, and Table~\ref{tab:fnorm} includes the average flux correction for our transit observations. 

\subsection{Stellar Variability Monitoring}
\label{sec:ait_obs}
We acquired 135 out-of-transit $R$-band images over the 2014-2015, 2015-2016, and 2016-2017 observing seasons with Tennessee State University's 14-inch Celestron Automatic Imaging Telescope (AIT). These observations, however, do not include the epochs of the \textit{Spitzer} and ground-based \citep{Chen17,Louden17} transit observations.  We therefore used 740 out-of-transit $V$-band images from Ohio State University's All-Sky Automated Survey for Supernovae\footnote{\url{https://asas-sn.osu.edu/}} (ASAS-SN) program \citep{Shappee14,Kochanek17} for more detailed activity monitoring coverage. The ASAS-SN observations were taken over the 2013$-$2014, 2014$-$2015, 2015$-$2016, and 2016$-$2017 observing seasons. The number of observations, date range, mean magnitude, and standard deviation of the data taken in each observing season are included in Table~\ref{tab:activity}. 

\subsection{Estimating Activity Levels}
\label{sec:activity_levels}
To quantify the level of activity in WASP-52, we used observations from the Advanced CCD Imaging Spectrometer (ACIS) on the \textit{Chandra} X-ray telescope. These data were taken from the \textit{Chandra} public archive (GO 15728; PI Wolk).  Standard CIAO\footnote{CIAO (Chandra Interactive Analysis of Observations) 4.9 and CALDB 4.7.4 versions were used in the analysis \citep{frus06}.} tasks were applied to reduce the data. The software ISIS \citep{isis} was used to fit the spectrum with a metallicity fixed to photospheric values ([Fe/H]=0.03) and the ISM absorption assumed to be N$_{\rm H}$ = 4 $\times$ 20 cm$^{-3}$ (based on target location and distance), consistent with the fit. The resulting 1-T fit has  log~T (K)= 6.54$^{+1.06}_{-0.29}$ and log EM (cm$^{-3}$) = 51.14$^{+0.33}_{-0.63}$. We measured an X-ray luminosity of $L_{\rm x}$ = ($3.5 \pm 1.0$) $\times 10^{28}$ erg s$^{-1}$ in the 0.12$-$2.48 keV band (Sanz-Forcada et al., in prep.). The light curve shows variability, but poor statistics hamper a
detailed analysis. The corresponding log($L_{\rm x}/L_{\rm bol}$) = $-$4.7 indicates that WASP-52 is a moderately active star. Further details will be included in Sanz-Forcada et al. (in prep.)      

\subsection{Modeling the Variability Monitoring Data}
\label{sec:GPs}
We initially adopted a simple sinusoidal model of period P = 11.8 $\pm$ 3.3 days based on the \citet{Hebrard13} rotational period, but found that this approach did not accurately model the variations in amplitude and period of the AIT and ASAS-SN variability monitoring data over all observing seasons. Discrepancies between the sinusoidal model and the photometric monitoring data are likely due to different spot configurations at different observed epochs \citep{Dang18}, suggesting that a quasi-periodic model would more accurately fit the data. 

\begin{figure*}
\centering
\includegraphics[scale=0.80]{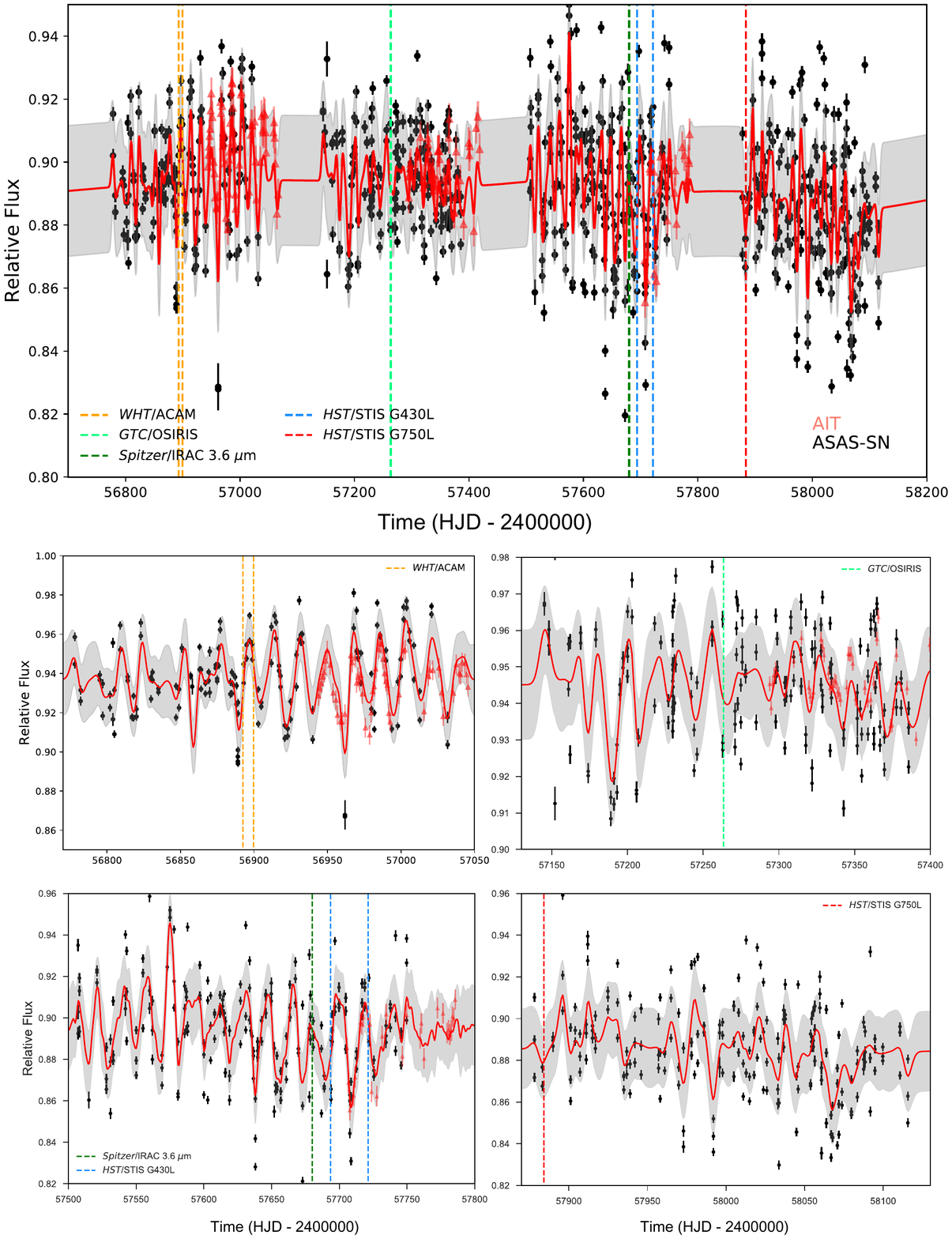}
\centering
\caption{Ground-based photometric observations of WASP-52 from ASAS-SN (black points) and AIT (red triangles) during the 2013$-$2014 (middle left), 2014$-$2015 (middle right), 2015$-$2016 (bottom left), and 2016$-$2017 (bottom right) observing seasons. The data are flux relative to the average brightness of comparison stars. The Gaussian process regression model (red) and 1$\sigma$ uncertainty (gray) fit to the combined ASAS-SN and AIT data are overplotted for the full dataset (top panel) and each observing season separately. The dashed vertical lines indicate the \textit{WHT}/ACAM (orange), \textit{GTC}/OSIRIS (light green), \textit{HST}/STIS G430L (blue), \textit{HST}/STIS G750L (red), and \textit{Spitzer}/IRAC (dark green) transit epochs.}
\label{fig:activity}
\end{figure*}

We therefore jointly modeled the AIT and ASAS-SN ground-based stellar activity data using a Gaussian process (GP) regression, a framework which has been shown to accurately disentangle stellar activity signals from planetary signals in radial velocity data (e.g., \citealt[]{Aigrain12,Haywood14}) and photometry (e.g., \citealt[]{Pont13,Aigrain16,Angus17}). 
We ran a GP optimization routine \citep{GeorgeCode} using the {\tt{George}} package for Python. We used a three component kernel to model the quasi-periodicity of the ground-based activity monitoring data, irregularities in the amplitude of the ground-based photometry, and stellar noise.
See Appendix~\ref{sec.kernel} for further details regarding the functional form of our chosen kernel. We use a gradient based optimization routine to find the best-fit hyperparameters and set the 11.8$\pm$3.3 day rotation period from \citet{Hebrard13} as a uniform prior with an uncertainty three times larger than the literature value. Figure \ref{fig:activity} shows the ground-based variability monitoring data overplotted with the Gaussian process model for all seasons modeled jointly and for each observing season separately.


The GP regression model with a quasi-periodic kernel reproduces the observed flux variations well for epochs during which we have stellar activity data, but has large uncertainties outside of a given season. The model accurately predicts the stellar activity behavior for each observing season when fit separately, but this model prediction has a significantly larger 1$\sigma$ uncertainty when fitting the data from all seasons together. 
We therefore modeled the data for each observing season separately, and used the amplitude of the photometric variation given by the GP model for each epoch to correct the transmission spectrum for the effects of stellar activity (see Section \ref{sec:corr}). 


\capstartfalse

\begin{deluxetable}{cccc}
\tabletypesize{\scriptsize}
\tablewidth{0pt}
\tablecolumns{4}
\tablecaption{Stellar flux values (normalized by the non-spotted stellar flux) for each transit observation \& average flux correction for each bandpass \label{tab:fnorm}}

\tablehead{\colhead{Instrument} & \colhead{$f_{norm}$} & \colhead{Error} & \colhead{$\Delta f$ } }

\startdata 
STIS (visit 52) & 0.978 & 0.009  & 0.039 \\
STIS (visit 53) & 0.955 & 0.016  & 0.039 \\
STIS (visit 54) & 0.972 & 0.016  & 0.022 \\
IRAC  			& 0.959 & 0.009  & 0.006 \\
\enddata

\end{deluxetable}
\label{tab:fnorm}
\capstarttrue

\subsection{Correcting for Unocculted Spots}
\label{sec:corr}
The temperature difference between the stellar photosphere and the spotted region introduces a slope in the planet spectrum (e.g., \citealt{Kreidberg17}), so it is necessary to correct for this effect.  Using the results of the GP regression model described in Section~\ref{sec:GPs}, we followed the prescription of \citet{Huitson13} to correct the spectrophotometric light curves for the effect of unocculted stellar spots of a fixed size. This method involves estimating the variability amplitude for a broadband wavelength value (determined by the photometric filter of the activity monitoring data), which can then be used as an anchor for the wavelength-dependent flux correction on the \textit{HST} and \textit{Spitzer} data.

We first converted the ASAS-SN and AIT photometric variability monitoring data to relative flux. Excluding in-transit measurements, we estimated the non-spotted stellar flux to be $F_{\star}$ = max($F$) + $k\sigma$, where $F$ is the variability monitoring data, $\sigma$ is the dispersion of the photometric measurements, and $k$ is a factor fixed to unity (\citealt{Aigrain12}; see also Appendix \ref{sec.corr}). The variability monitoring data was then normalized to the non-spotted stellar flux to estimate the amount of dimming. Table \ref{tab:fnorm} gives the dimming values for each transit observation. We also computed the amplitude of the spot corrections at the variability monitoring wavelength $\Delta f_{0}$ = $1 - f_{norm}$, where $f_{norm}$ is the mean of the normalized flux array. 

\begin{figure*}
\centering
\includegraphics[scale=0.75]{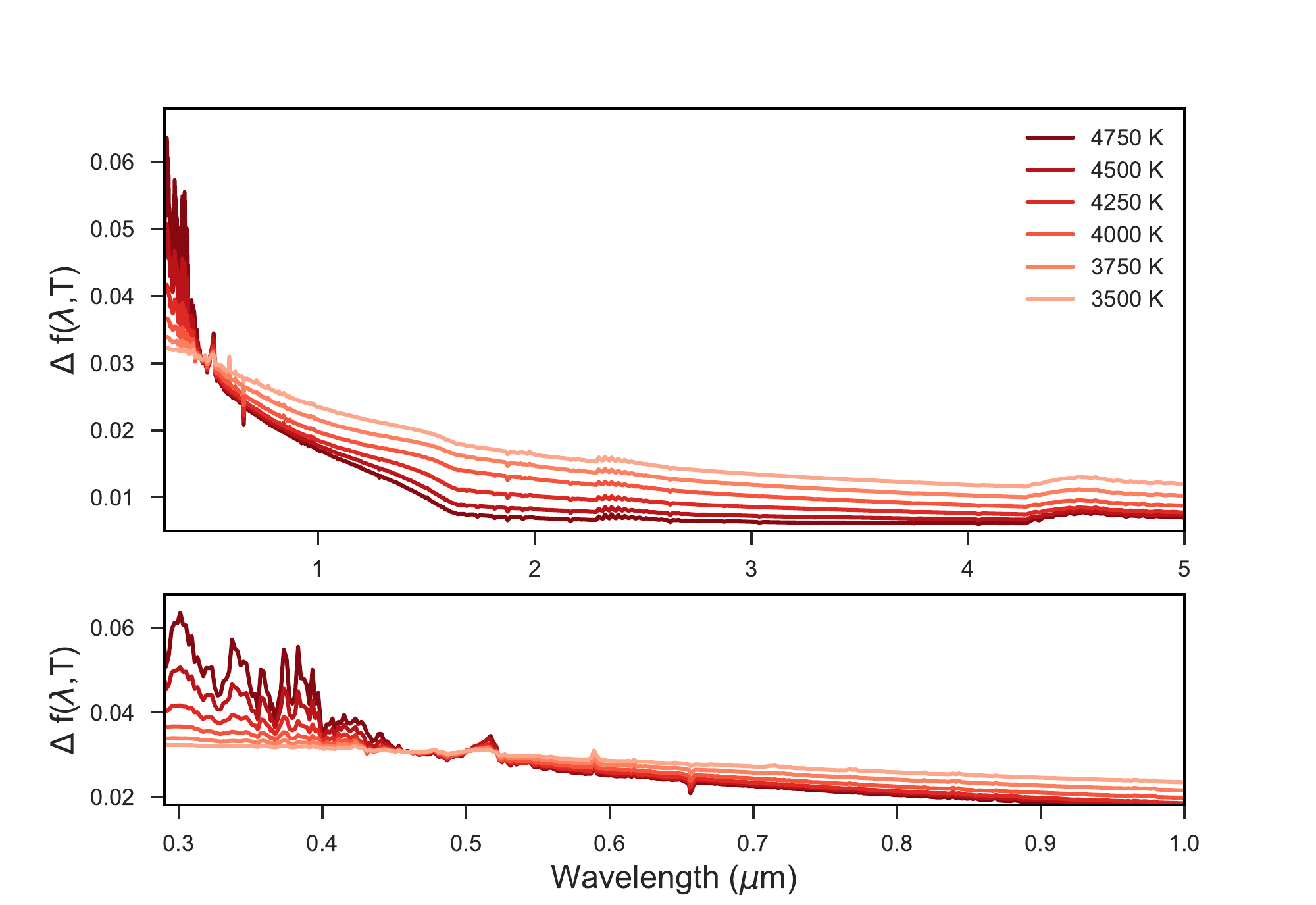}
\centering
\caption{Theoretical dimming of unocculted spots in the wavelength range of the \textit{HST}/STIS and \textit{Spitzer}/IRAC observations for stellar flux models ranging in temperature from 3500 $-$ 4750 K (corresponding to temperature differences between the spot and stellar photosphere from 1500 K to 250 K). The dimming $\Delta f(\lambda,T)$ is derived by multiplying the wavelength-dependent flux correction by the non-spotted stellar flux (Section \ref{sec:corr}). \textit{Top panel}: Spot dimming for the full STIS+\textit{Spitzer} wavelength range ($\sim$0.29-4.5 $\mu$m). \textit{Bottom panel}: Zoom-in of the STIS wavelength range only ($\sim$0.29-1.0 $\mu$m).}
\label{fig:spot_dimming}
\end{figure*}

To derive the wavelength-dependent flux correction, we used a stellar flux model ($T_{eff}$ = 5000 K, log($Z$) = $-$1.5, log($g$) = 4.5) and a spot model ($T_{eff}$ = 4750 K, log($Z$) = $-$1.5, log($g$) = 4.5) from the \citet{Kurucz93} 1D {\tt{ATLAS}} grid\footnote{\url{http://kurucz.harvard.edu/}} of stellar atmospheric models. The stellar flux model was chosen based on the effective temperature, metallicity, and surface gravity of WASP-52 given in \citet{Hebrard13}. The spot model is the same as the stellar model but 250 K cooler \citep{Berdyugina05}. 
To select this spot model, we tested different cold spots ranging from 3500$-$4750 K (corresponding to temperature differences between the spot and stellar photosphere from 1500 K to 250 K). Figure \ref{fig:spot_dimming} shows that spots at colder temperatures exhibit less flux dimming at longer wavelengths and a weaker slope in the optical. To correct for the effect of unocculted spots, we therefore use a spot model at 4750 K for which cold spots give the strongest slope to account for the maximum possible contribution from star spots in the data.

We interpolated the stellar model to the spot model grid and computed the wavelength-dependent correction factor derived in \citet{Sing11}:

\begin{equation}
f(\lambda,T) = \Big( 1 - \frac{F_{\lambda,T_{spot}}}{F_{\lambda,T_{star}}} \Big) \bigg/ \Big( 1 - \frac{F_{\lambda_{o},T_{spot}}}{F_{\lambda_{o},T_{star}}}\Big)
\label{eqn:flux_corr}
\end{equation}
where $F_{\lambda,T_{spot}}$ is the stellar model flux at temperature $T_{spot}$ and at the wavelength of the transit observations, $F_{\lambda,T_{star}}$ is the stellar model flux at the wavelength of the transit observations and at temperature $T_{star}$, $F_{\lambda_{o},T_{spot}}$ is the stellar model flux at temperature $T_{spot}$ at the reference wavelength of the activity monitoring data, and $F_{\lambda_{o},T_{star}}$ is the stellar model flux at temperature $T_{star}$ at the activity monitoring reference wavelength. The final flux dimming correction was then calculated as $\Delta f$ = $\Delta f_{0}$ $\times$ $f(\lambda,T)$.

We computed the average flux correction over each bandpass (see Table \ref{tab:fnorm}) and applied the correction to each spectrophotometric light curve using: 

\begin{equation}
y_{corrected} = y + \frac{\Delta f}{(1 - \Delta f)} \overline{y_{oot}}
\end{equation}
where $y_{corrected}$ is the corrected light curve flux, $y$ is the original (uncorrected) light curve, and $\overline{y_{oot}}$ is the mean of the out of transit exposures. We then fit analytic transit light curve models \citep{Mandel02} to the stellar activity corrected light curves, as detailed in Section \ref{sec:analysis}. 

\section{Light Curve Fits}
\label{sec:analysis}
We extracted the broadband transmission spectrum and fit the spectrophotometric light curves following the methods described in \citet{Sing11,Sing13} and \citet{Nikolov14}, and they are briefly summarized here. To simultaneously fit for the transit and systematic effects, we modeled each of the STIS and \textit{Spitzer} transit light curves with a two-component function consisting of a transit model multiplied by a systematics model.
We adopted the complete analytic transit models of \cite{Mandel02}, which are parametrized by the mid-transit time $T_{0}$, orbital period $P$, inclination $i$, normalized planet semi-major axis $a/ R_{\star}$, and planet-to-star radius ratio $R_{{\rm{p}}} /R_{\star}$. 

\capstartfalse

\begin{deluxetable*}{ccccc}
\tabletypesize{\scriptsize}
\tablewidth{0pt}
\tablecolumns{5}
\tablecaption{Derived System Parameters for WASP-52b\label{tab:params}}

\tablehead{\colhead{}  & \colhead{\citet{Hebrard13}} & \colhead{\citet{Chen17}} & \colhead{\citet{Louden17}} & {This work\tablenotemark{a}}  }
\startdata 
Period~ $P$ [days]                     &   1.7497798 $\pm$ 0.0000012            &  1.7497798 (fixed)     &  1.74978089 $\pm$ 0.00000013   & 1.749779800 (fixed)     \\
Orbital inclination~ $i$ [$^{\circ}$]  &   85.35 $\pm$ 0.20                     &  85.06 $\pm$ 0.27      &  85.33 $\pm$ 0.22     	      & 85.17 $\pm$ 0.13        \\
Orbital eccentricity~ $e$              &   0.0 (fixed)                          &  0.0 (fixed)       	 &  0.0 (fixed)                   & 0.0 (fixed)             \\
Scaled semi-major axis~ $a/R_{\star}$  &   7.38 $\pm$ 0.10                      &  7.14 $\pm$ 0.12    	 &  7.23 $\pm$ 0.12      	      & 7.22 $\pm$ 0.07         \\
Radius ratio~ $R_{p}/R_{\star}$ 	   &   0.1646 $\pm$ 0.0020                  & 0.1608 $\pm$ 0.0018 	 &  0.1741 $\pm$ 0.0063  	      & 0.1639 $\pm$ 0.0005     \\
\enddata
%
\tablenotetext{a}{The values reported here are the weighted mean of fitted system parameters from the \textit{Spitzer} observations.}
\end{deluxetable*}

\label{tab:params}
\capstarttrue

\subsection{STIS White Light Curves}
\label{sec:stis_wlc}

\begin{figure*}
\centering
\includegraphics[scale=0.80]{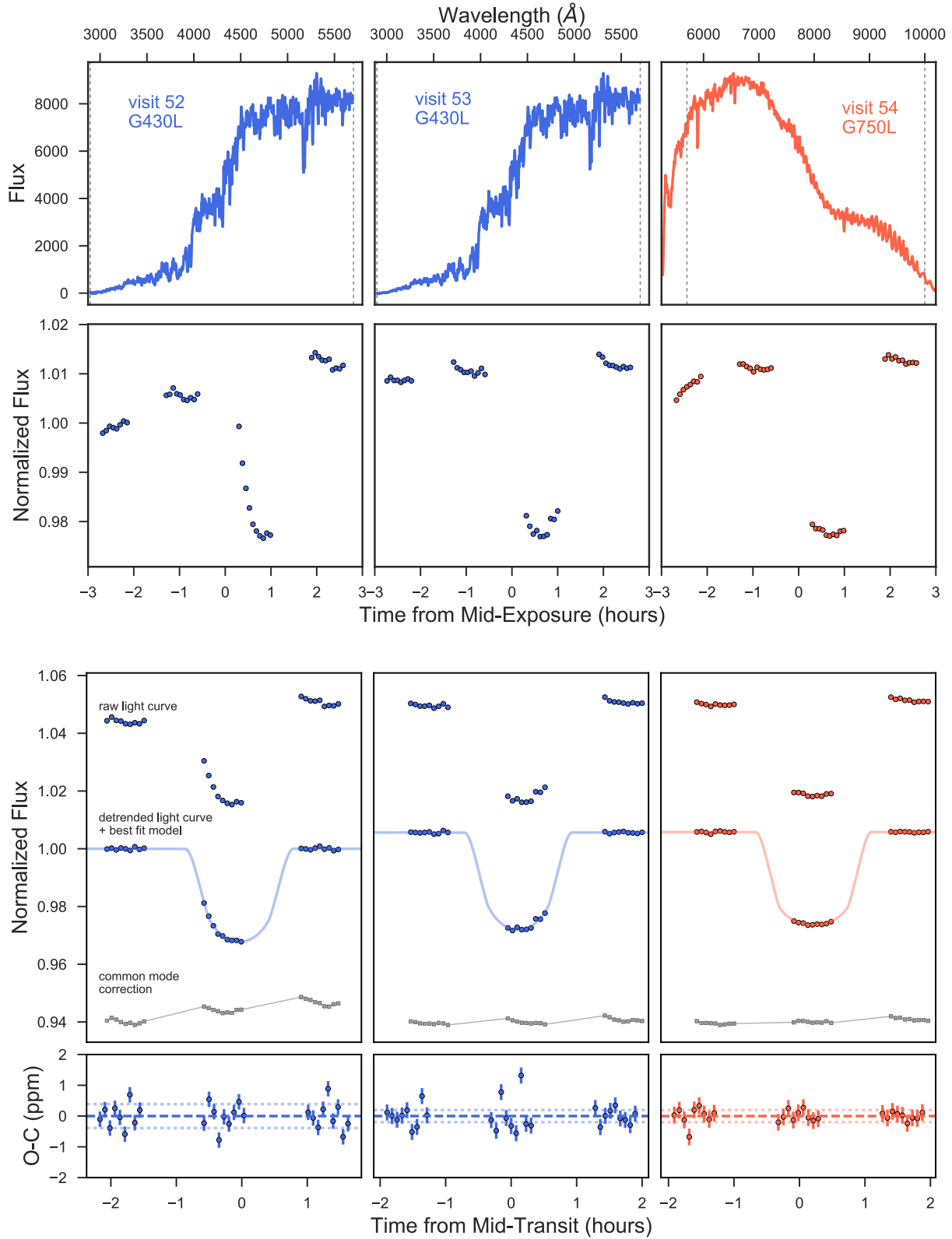}
\centering
\caption{\textit{HST}/STIS stellar spectra of WASP-52b and the corresponding white light curves for visits 52 (left column), 53 (middle column), and 54 (right column). \textit{Top panel}: Example stellar spectra taken with the G430L (blue) and G750L (red) grating. The vertical gray dashed lines indicate the wavelength range used to produce the white light curves. \textit{Second panel}: The raw white light curves for each visit. \textit{Third panel}: The raw and detrended light curves (excluding the first orbit and the first exposure of each subsequent orbit) with the best-fit model overplotted. The common mode correction (gray squares) is the transit+systematics model divided by the best-fit transit model. \textit{Bottom panel}: Transit fit residuals with error bars.}
\label{fig:wlc}
\end{figure*}

We produced the STIS broadband (wavelength-integrated) light curves by summing the time series over the complete wavelength range of the bandpass (2892$-$5700~\AA~ for the G430L grating; 5240$-$10270~\AA~ for the G750L grating). The white light curves for each of the STIS visits are shown in Figure \ref{fig:wlc}. Photometric uncertainties were derived based on pure photon statistics. The raw white light curves exhibited instrumental systematics related to the orbital motion of the spacecraft (\citealt{Gilliland99,Brown01}). In particular, the \textit{HST} focus is known to experience significant variations on the spacecraft orbital time-scale resulting from thermal expansion/contraction during the spacecraft's day/night orbital cycle. 

As in past studies (e.g., \citealt{Huitson13,Sing13,Nikolov14}), we accounted for and detrended these instrumental systematic effects by fitting a fourth-order polynomial to the flux dependence on \textit{HST} orbital phase. We excluded the first orbit and the first exposure of each subsequent orbit in accordance with common practice, since these data have unique, complex systematics (see Figure \ref{fig:wlc}) and were taken while the telescope was thermally relaxing into its new pointing position. We applied orbit-to-orbit flux corrections to the STIS data by fitting a polynomial of the spacecraft orbital phase ($\phi_{\rm{t}}$), drift of the spectra on the detector ($x$ and $y$), the shift of each stellar spectrum cross-correlated with the first spectrum of the time series ($\omega$), and time ($t$). 

We then generated systematics models spanning all possible combinations of detrending variables (see Appendix \ref{sec.sys} and Table \ref{tab:systematics} for details), and performed separate fits using each systematics model included in the two-component function. For each model, we fixed $P$, $i$, and $a/R_{\star}$ to the values given in \citet{Hebrard13}, assumed zero eccentricity, and fit for $T_{0}$, $R_{p}/R_{\star}$, stellar baseline flux, and instrument systematic trends. We determined the best fit parameters of the two-component model using a Levenberg-Marquardt least-squares fitting routine \citep{Markwardt09} to derive the system parameters and calculated the Akaike Information Criterion (AIC; \citealt{Akaike74}) for each function. The results of these fits are shown in Table \ref{tab:params}. The STIS stellar spectra and raw and detrended white light curves for each visit are shown in Figure \ref{fig:wlc}.

We marginalized over the entire set of functions following the framework outlined in \citet{Gibson14}. Marginalization over multiple systematics models assumes equally weighted priors for each model tested. Table \ref{tab:systematic_models} shows the results for each systematics model using the STIS white light curves. The reduced chi-squared $\chi_{r}$ from the fits can be used as a proxy for the photon noise level of a given data set \citep{Nikolov14}. We selected the systematics model for use in detrending the STIS white light curves based on the lowest AIC value, since the light curves show no significant correlation to the quantified systematic parameters \citep{Nikolov14}.


\subsection{STIS Spectrophotometric Light Curves}
\label{sec:lc_bins}

We produced the spectrophotometric light curves by dividing the spectra into 17 $-$ 400 \AA~width bins and integrating the flux from each bandpass. The width of the bins is determined primarily by the need to achieve a given photometric precision to be sensitive to features in the planet's transmission spectrum that are comparable in amplitude to an atmospheric scale height. The smaller bin sizes were chosen to be centered at specific absorption features, such as Na {\sc i} at 5893 \AA~and K {\sc i} at 7665 \AA. 

We modeled the systematic errors using two methods. In the first approach, we independently fit each of the binned light curves with the same family of transit+systematics models (see Appendix \ref{sec.sys}) as the broadband light curve. The only differences are that we fixed the mid-transit time $T_{0}$ and the scaled semi-major axis $a/R_{\star}$ to the white light curve best fit values. We also fixed the  limb darkening coefficients (computed following the procedure outlined in Section \ref{sec:ld}) to the derived theoretical values. In the second approach, we performed a common mode correction to remove color-independent systematic trends from each spectral bin. Then, we fit the common mode corrected spectroscopic light curves by fitting for residuals with a parametrized model of six fewer free parameters ($c_{1}-c_{4}$, $T_{0}$, and $a/R_{\star}$) and marginalizing over the entire set of functions defined in Appendix \ref{sec.sys}. The common mode trends are computed by dividing the raw flux of the white light curve in each grating by the best fit analytic transit model. We applied the common mode correction by dividing each binned light curve by the derived common mode flux. Removing common mode trends is known to reduce the amplitude of the observed \textit{HST} breathing systematics, since these trends are similar in wavelength across the detector \citep{Sing13,Nikolov14}. The common mode corrected light curves are shown in Figure \ref{fig:wlc}.

\begin{figure*}
\centering
\includegraphics[scale=0.82]{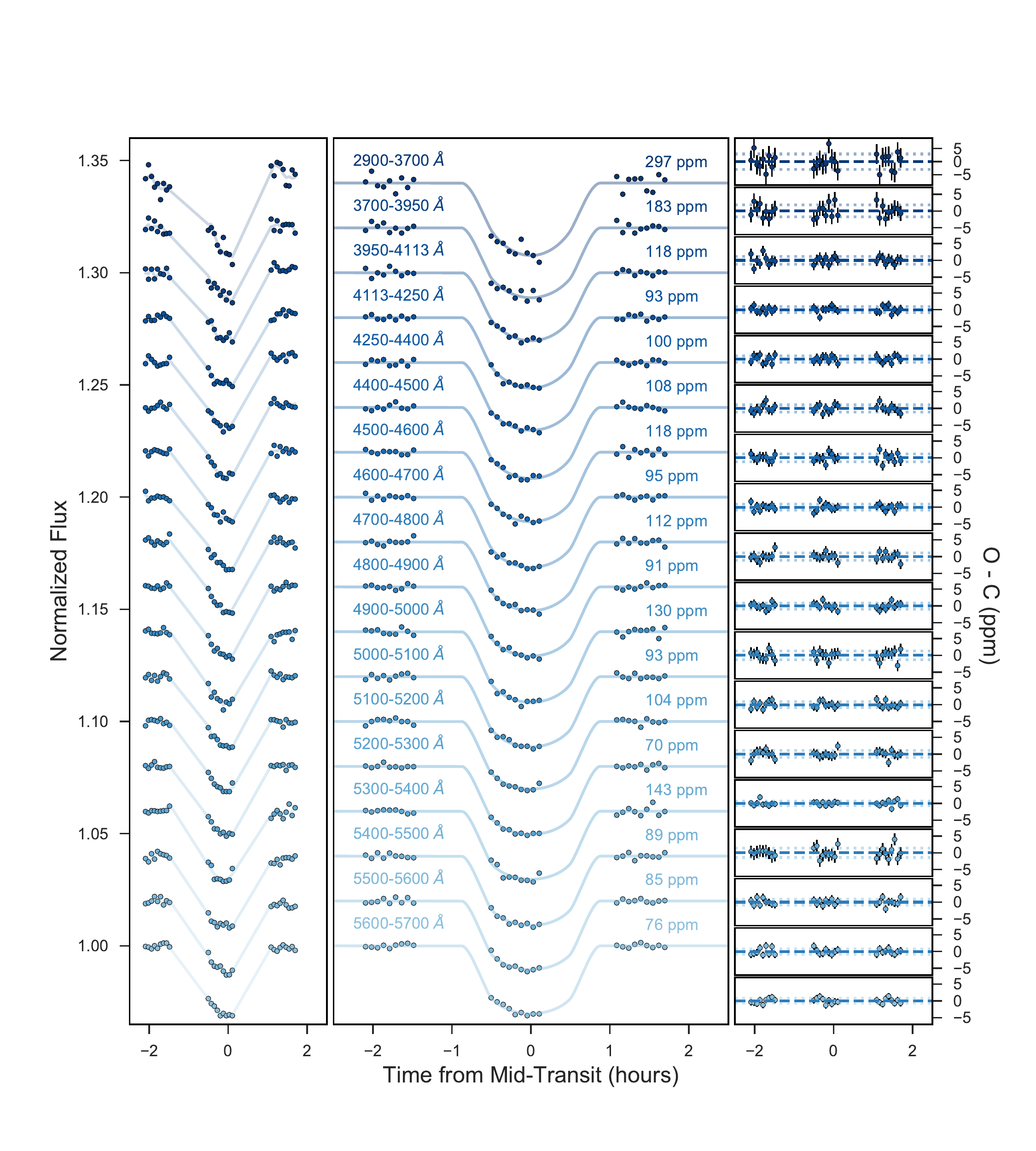}
\centering
\caption{\textit{HST}/STIS G430L observations of WASP-52b visit 52, excluding the first orbit and the first exposure of each subsequent orbit. Raw (left panel) and detrended (middle panel) light curves are shown for each wavelength bin, and are offset vertically by an arbitrary constant for clarity. Observed minus computed residuals with error bars are shown in the right panel.}
\label{fig:binfits_G430L_1}
\end{figure*}

\begin{figure*}
\centering
\includegraphics[scale=0.82]{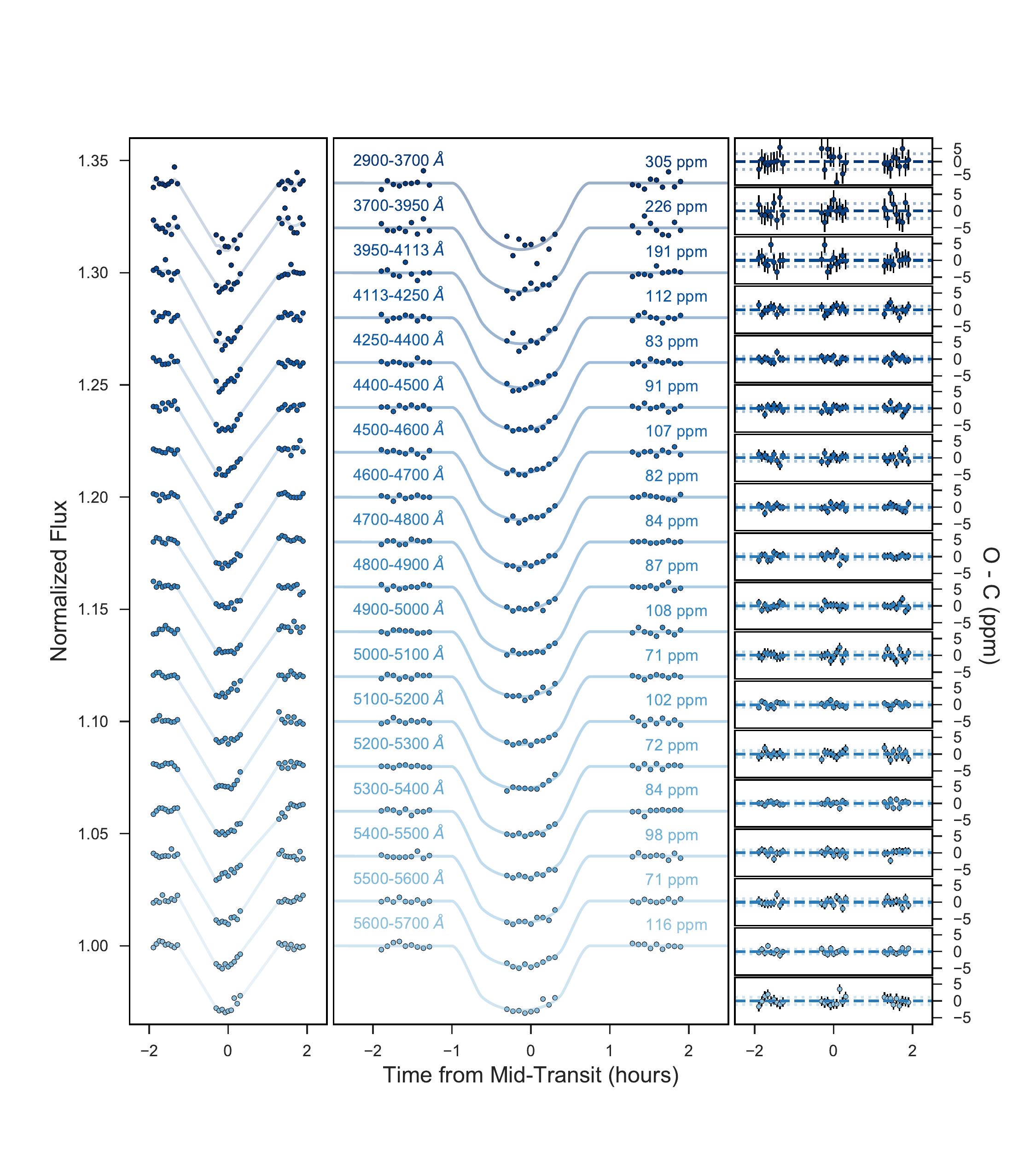}
\centering
\caption{Same as Figure \ref{fig:binfits_G430L_1}, but for visit 53.}
\label{fig:binfits_G430L_2}
\end{figure*}

\begin{figure*}
\centering
\includegraphics[scale=0.82]{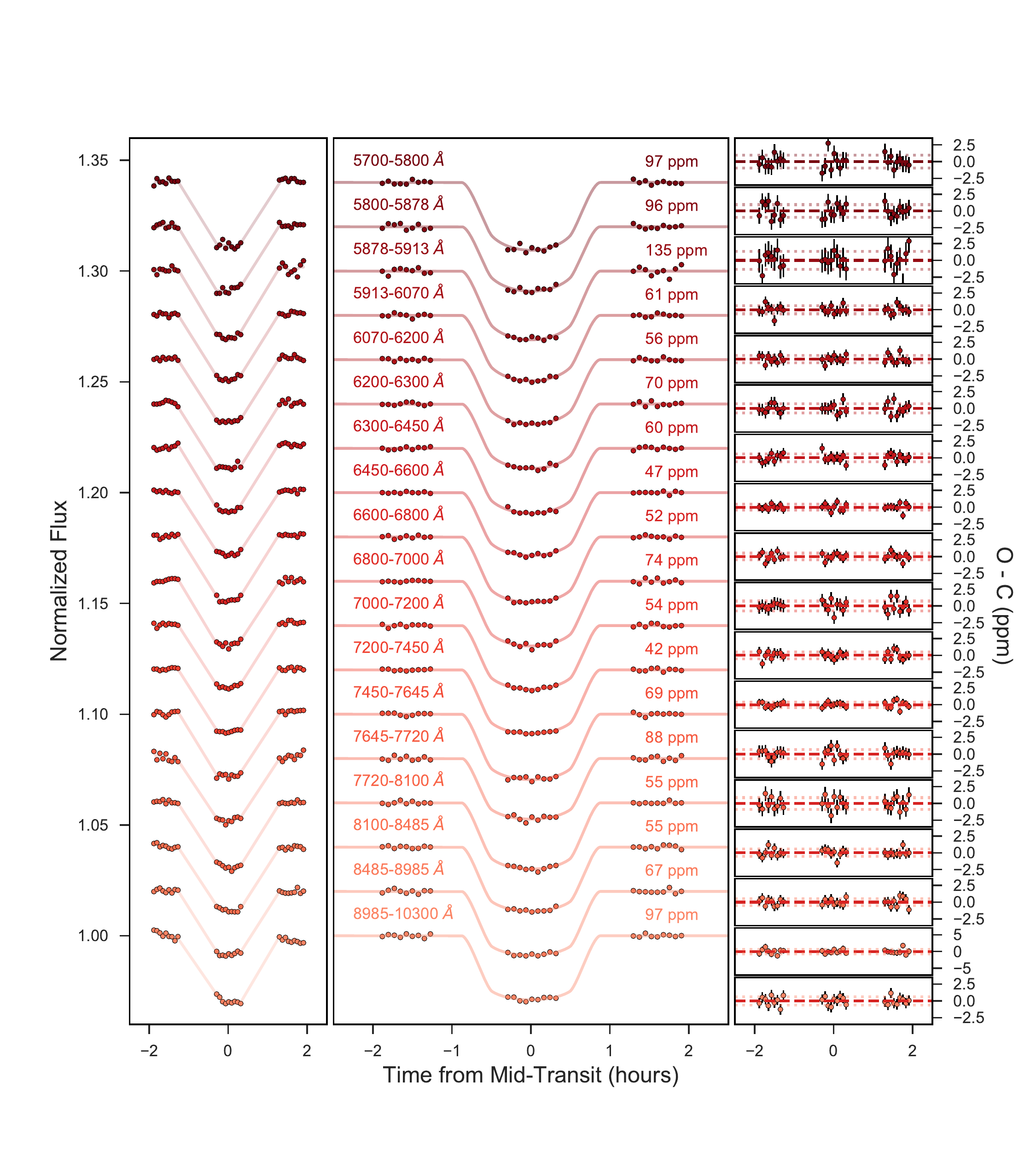}
\centering
\caption{\textit{HST}/STIS G750L observations of WASP-52b visit 54, excluding the first orbit and the first exposure of each subsequent orbit. Raw (left panel) and detrended (middle panel) light curves are shown for each wavelength bin, and are offset vertically by an arbitrary constant for clarity. Observed minus computed residuals with error bars are shown in the right panel.}
\label{fig:binfits_G750L}
\end{figure*}

Both methods produced similar results (i.e., consistent baseline in $R_{p}/R_{\star}$). Since the common mode correction produces lower dispersion in the spectrophotometric light curves and smaller $R_{p}/R_{\star}$ uncertainties \citep{Nikolov15}, we report the common mode corrected results for the fitted $R_{p}/R_{\star}$ (before and after applying the correction for unocculted spots discussed in Section \ref{sec:corr}) and non-linear limb darkening coefficients for each spectroscopic channel in Table \ref{tab:tr_spec}. The raw and detrended STIS spectrophotometric light curves for the G430L and G750L gratings are shown in Figures \ref{fig:binfits_G430L_1}, \ref{fig:binfits_G430L_2}, and \ref{fig:binfits_G750L}. 

\subsection{IRAC Light Curves}
\begin{figure*}
\centering
\includegraphics[scale=0.85]{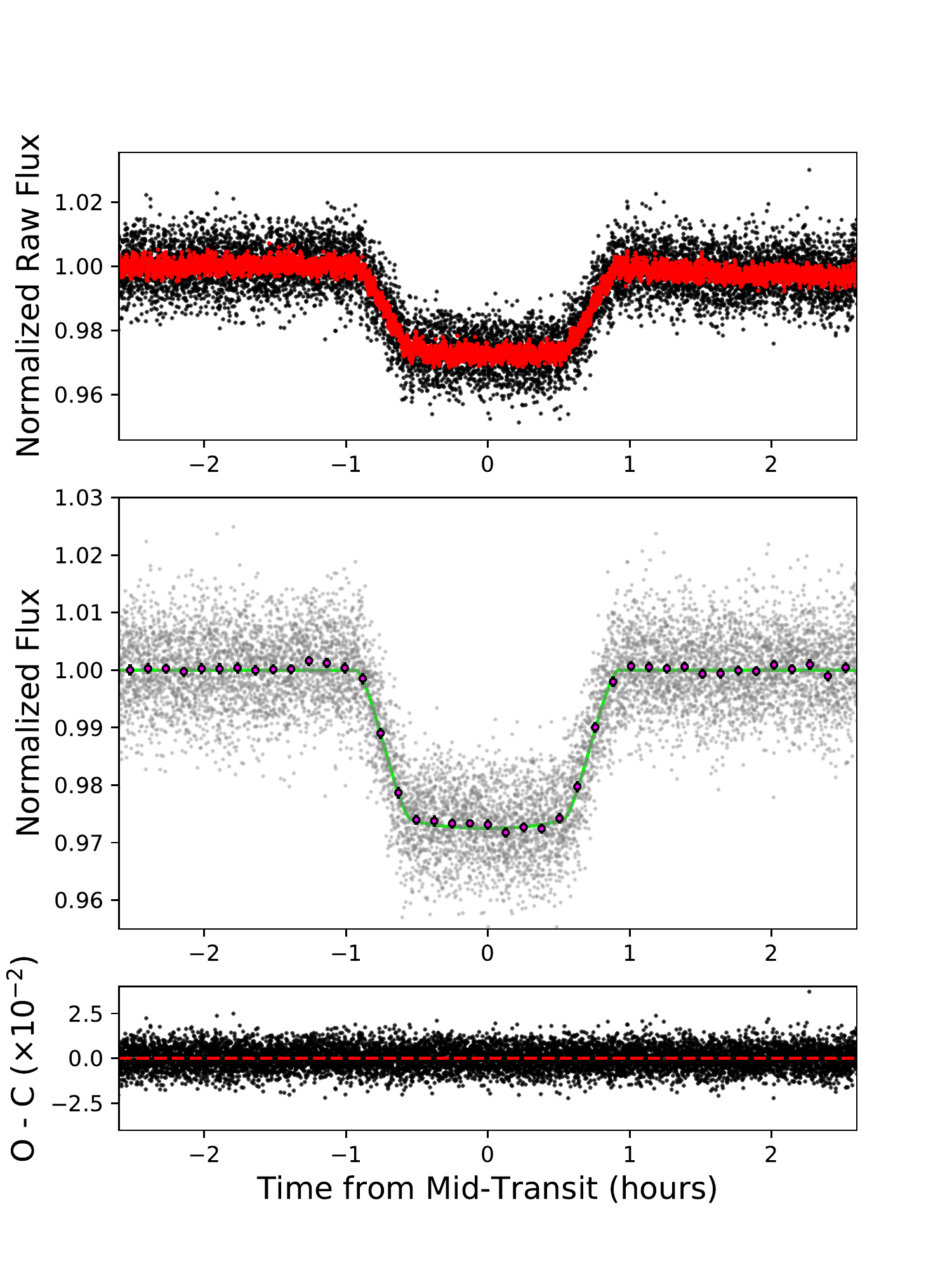}
\centering
\caption{\textit{Spitzer}/IRAC 3.6 $\mu$m transit light curve. \textit{Top panel}: Raw flux (black points) and best-fit transit model (red points). \textit{Middle panel}: Detrended light curve (gray points) and best-fit transit model (green line) overlaid with the binned light curve (magenta points; 45 bins of 224 data points each). \textit{Bottom panel}: observed minus computed residuals  (black points) of the raw light curve.}
\label{fig:spitzer_lc}
\end{figure*}

\begin{figure*}
\centering
\includegraphics[scale=0.85]{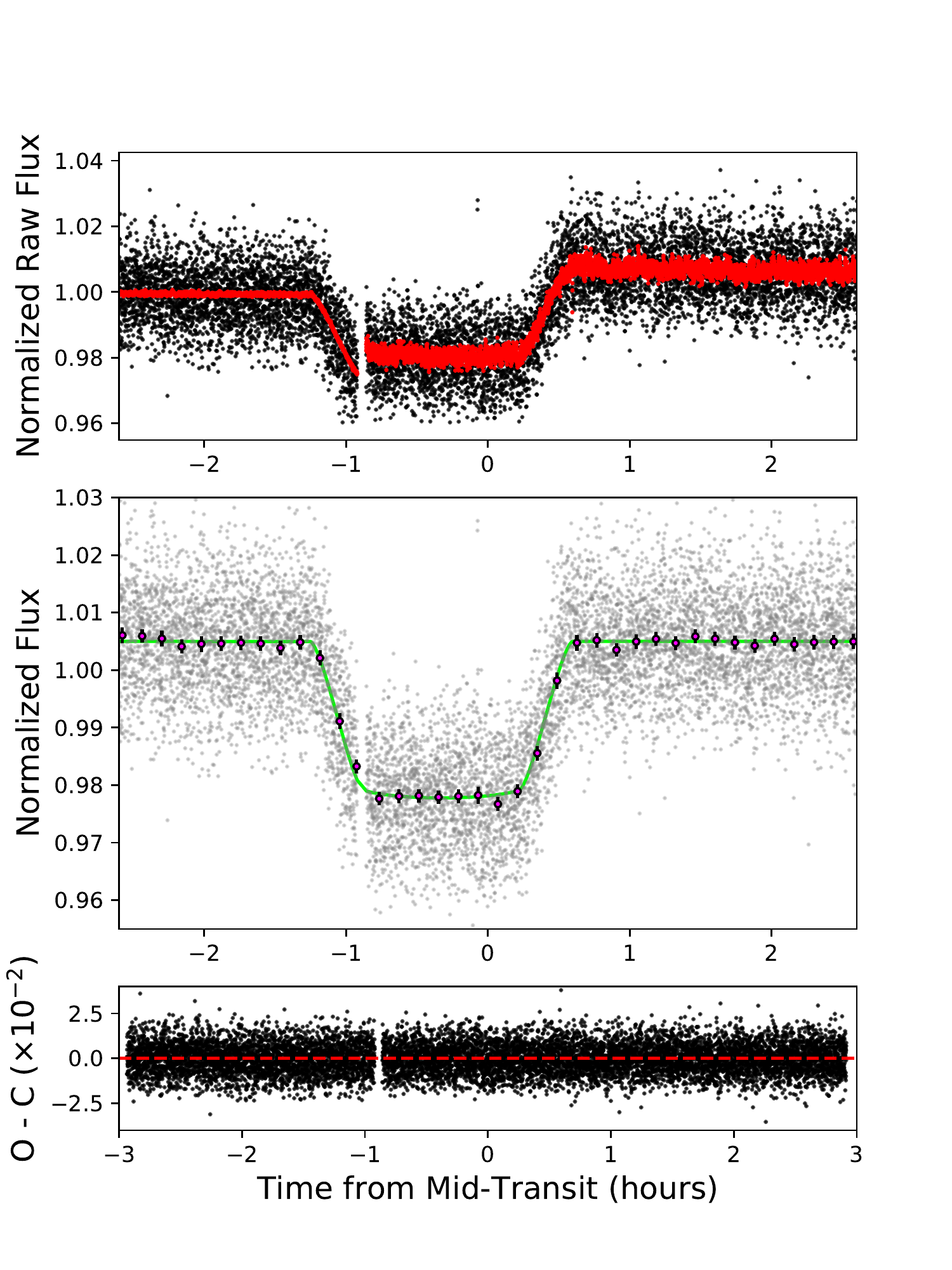}
\centering
\caption{\textit{Spitzer}/IRAC 4.5 $\mu$m transit light curve. \textit{Top panel}: Raw flux (black points) and best-fit transit model (red points). The small gap at the end of the transit ingress corresponds to the delay between subsequent data readouts. \textit{Middle panel}: Detrended light curve (gray points) and best-fit transit model (green line) overlaid with the binned light curve (magenta points; 43 bins of 239 data points each). \textit{Bottom panel}: observed minus computed residuals  (black points) of the raw light curve.}
\label{fig:spitzer_lc_45}
\end{figure*}

We modeled the 3.6 $\mu$m and 4.5 $\mu$m IRAC transit photometry in accordance with the methods outlined in \citet{Sing11,Sing13} and \citet{Nikolov14}.
To correct for flux variations from intrapixel sensitivity, we fit a polynomial to the stellar centroid position (\citealt{Reach05,Charbonneau05,Charbonneau08,Knutson08}). This technique is effective on short timescales ($<$10 hours) and for small ($<$0.2 pixels) variations in the stellar centroid position \citep{Lewis13}. We corrected for systematic effects using a model given by the linear combination: 

\begin{equation}
f(t) = a_{0}~ + a_{1}x~ + a_{2}x^{2}~ + a_{3}y~ + a_{4}y^{2}~ + a_{5}xy~ + a_{6}t
\label{eq:irac}
\end{equation}
where $f(t)$ is the stellar flux as a function of time, the coefficients $a_{0}$ through $a_{6}$ are free fitting parameters, $x$ and $y$ are the detector positions of the stellar centroid, and $t$ is time. We generate all possible model combinations of Equation \ref{eq:irac}, which we marginalize over using the \citet{Gibson14} procedure as detailed in Section \ref{sec:stis_wlc} and Appendix \ref{sec.sys}. Using the WASP-52 system parameters from \citet{Hebrard13} as priors, we jointly fit for all parameters and find that our results (see Table \ref{tab:params}) agree within 1$\sigma$ with the STIS white light curve analysis. To measure $R_{p}/R_{\star}$ for the transmission spectrum, we fixed the orbital period $P$, normalized planet semi-major axis $a/R_{\star}$, inclination $i$, and central transit time $T_{0}$ to the best fit values from the joint fit. The limb darkening coefficients were also fixed to their theoretical values based on 3D stellar atmosphere models (see Section \ref{sec:ld}). The measured planetary radius and limb darkening coefficients are included in Table \ref{tab:tr_spec}. Figures \ref{fig:spitzer_lc} and \ref{fig:spitzer_lc_45} show the raw and detrended \textit{Spitzer} transit light curves. 


\subsection{Limb Darkening Models}
\label{sec:ld}
We modeled the limb darkening of WASP-52b using the four parameter non-linear limb darkening law \citep{Claret00} given by:

\begin{equation}
\frac{I(\mu)}{I(1)} = 1 ~-~ \sum_{n=1}^{4}c_{n}(1-\mu^{n/2})
\end{equation}
where $I$(1) is the intensity at the center of the stellar disk, $c_{n}$ ($n$ = 1$-$4) are the limb darkening coefficients, and $\mu$ = cos($\theta$), where $\theta$ is the angle between the normal to the stellar surface and the line of sight.   

To derive the stellar limb darkening coefficients, we followed the procedure described in \citet{Sing10} and initially used values for the four limb darkening coefficients based on 1D {\tt{ATLAS}} theoretical stellar models \citep{Kurucz93}. We then derived the limb darkening coefficients from 3D stellar models \citep{Magic15} and compared to the 1D results to eliminate the known wavelength-dependent degeneracy of limb darkening with transit depth \citep{Sing08a}. This approach reduces the number of free parameters in the fit (typically four parameters per grating), but may cause an underestimation of errors in the derived spectrum. The derived non-linear 3D limb darkening coefficients for each spectrophotometric light curve are shown in Table \ref{tab:tr_spec}.

\capstartfalse

\begin{deluxetable*}{ccccccc}
\tabletypesize{\scriptsize}
\tablewidth{0pt}
\tablecolumns{7}
\tablecaption{Broadband transmission spectrum results for WASP-52b for the STIS G430L \& G750L and \textit{Spitzer} IRAC data
\label{tab:trspec}}

\tablehead{\colhead{$\lambda$ (\AA)}  & \colhead{$(R_{p}/R_{*})_{uncorr}$} & \colhead{$(R_{p}/R_{*})_{corr}$} & \colhead{$c_{1}$}  & \colhead{$c_{2}$}  & \colhead{$c_{3}$}  & \colhead{$c_{4}$} }

\startdata 
2900$-$3700   &  0.16957  $\pm$  0.00385  & 0.15421 $\pm$ 0.00677 & 0.4371 & -0.7679 & 1.6319 & -0.3645 \\ 
3700$-$3950   &  0.16684  $\pm$  0.00201  & 0.15781 $\pm$ 0.00296 & 0.7648 & -1.1573 & 1.7190 & -0.4157 \\
3950$-$4113   &  0.16703  $\pm$  0.00157  & 0.16681 $\pm$ 0.00296 & 0.4778 & -0.6459 & 1.6564 & -0.5511 \\
4113$-$4250   &  0.16871  $\pm$  0.00102  & 0.16457 $\pm$ 0.00159 & 0.4831 & -0.6750 & 1.7025 & -0.5879 \\
4250$-$4400   &  0.16672  $\pm$  0.00116  & 0.16394 $\pm$ 0.00148 & 0.6151 & -0.8347 & 1.7798 & -0.6519 \\
4400$-$4500   &  0.16618  $\pm$  0.00112  & 0.16101 $\pm$ 0.00142 & 0.4691 & -0.4858 & 1.5912 & -0.6570 \\
4500$-$4600   &  0.16685  $\pm$  0.00113  & 0.16358 $\pm$ 0.00149 & 0.4777 & -0.4094 & 1.5029 & -0.6494 \\
4600$-$4700   &  0.16827  $\pm$  0.00089  & 0.16583 $\pm$ 0.00106 & 0.5977 & -0.6932 & 1.6924 & -0.6811 \\
4700$-$4800   &  0.16625  $\pm$  0.00120  & 0.16211 $\pm$ 0.00187 & 0.4411 & -0.2389 & 1.1548 & -0.4505 \\
4800$-$4900   &  0.16630  $\pm$  0.00088  & 0.16290 $\pm$ 0.00115 & 0.5159 & -0.3136 & 1.2538 & -0.5560 \\
4900$-$5000   &  0.16557  $\pm$  0.00119  & 0.16079 $\pm$ 0.00156 & 0.4399 & -0.1314 & 1.0138 & -0.4335 \\
5000$-$5100   &  0.16806  $\pm$  0.00085  & 0.16463 $\pm$ 0.00105 & 0.5409 & -0.3989 & 1.1899 & -0.4566 \\
5100$-$5200   &  0.16888  $\pm$  0.00120  & 0.16623 $\pm$ 0.00175 & 0.5605 & -0.4363 & 1.0910 & -0.3757 \\
5200$-$5300   &  0.16769  $\pm$  0.00074  & 0.16541 $\pm$ 0.00103 & 0.5381 & -0.2913 & 1.0934 & -0.4736 \\
5300$-$5400   &  0.16589  $\pm$  0.00101  & 0.16330 $\pm$ 0.00114 & 0.5732 & -0.3581 & 1.1751 & -0.5264 \\
5400$-$5500   &  0.16802  $\pm$  0.00091  & 0.16416 $\pm$ 0.00132 & 0.5975 & -0.4018 & 1.1388 & -0.4758 \\
5500$-$5600   &  0.16665  $\pm$  0.00083  & 0.16279 $\pm$ 0.00108 & 0.6370 & -0.4886 & 1.2206 & -0.5147 \\
5600$-$5700   &  0.16621  $\pm$  0.00086  & 0.16157 $\pm$ 0.00154 & 0.5662 & -0.2476 & 0.9413 & -0.4076 \\
5700$-$5800   &  0.16782  $\pm$  0.00172  & 0.16566 $\pm$ 0.00170 & 0.5828 & -0.2962 & 1.0423 & -0.4802 \\ 
5800$-$5878   &  0.16556  $\pm$  0.00195  & 0.16346 $\pm$ 0.00192 & 0.6127 & -0.3245 & 0.9821 & -0.4278 \\
5878$-$5913   &  0.17087  $\pm$  0.00209  & 0.16858 $\pm$ 0.00206 & 0.6511 & -0.5383 & 1.2452 & -0.5387 \\
5913$-$6070   &  0.16643  $\pm$  0.00112  & 0.16434 $\pm$ 0.00110 & 0.6208 & -0.3332 & 0.9989 & -0.4502 \\
6070$-$6200   &  0.16515  $\pm$  0.00105  & 0.16310 $\pm$ 0.00105 & 0.6192 & -0.3316 & 0.9650 & -0.4310 \\
6200$-$6300   &  0.16611  $\pm$  0.00160  & 0.16407 $\pm$ 0.00157 & 0.6439 & -0.3583 & 0.9653 & -0.4321 \\
6300$-$6450   &  0.16590  $\pm$  0.00095  & 0.16388 $\pm$ 0.00094 & 0.6508 & -0.3875 & 0.9927 & -0.4467 \\
6450$-$6600   &  0.16371  $\pm$  0.00087  & 0.16179 $\pm$ 0.00086 & 0.6569 & -0.3766 & 0.9577 & -0.4418 \\
6600$-$6800   &  0.16633  $\pm$  0.00103  & 0.16439 $\pm$ 0.00102 & 0.6561 & -0.3747 & 0.9214 & -0.4106 \\
6800$-$7000   &  0.16419  $\pm$  0.00170  & 0.16231 $\pm$ 0.00168 & 0.6494 & -0.3485 & 0.8652 & -0.3871 \\
7000$-$7200   &  0.16491  $\pm$  0.00091  & 0.16307 $\pm$ 0.00090 & 0.6824 & -0.4389 & 0.9298 & -0.4051 \\
7200$-$7450   &  0.16448  $\pm$  0.00068  & 0.16268 $\pm$ 0.00067 & 0.6914 & -0.4691 & 0.9457 & -0.4158 \\
7450$-$7645   &  0.16656  $\pm$  0.00126  & 0.16479 $\pm$ 0.00126 & 0.6972 & -0.4678 & 0.9219 & -0.4080 \\
7645$-$7720   &  0.16113  $\pm$  0.00312  & 0.15943 $\pm$ 0.00309 & 0.7151 & -0.5133 & 0.9350 & -0.4061 \\
7720$-$8100   &  0.16805  $\pm$  0.00121  & 0.16632 $\pm$ 0.00121 & 0.6976 & -0.4689 & 0.8838 & -0.3862 \\
8100$-$8485   &  0.16538  $\pm$  0.00081  & 0.16372 $\pm$ 0.00080 & 0.7008 & -0.4971 & 0.8888 & -0.3849 \\
8485$-$8985   &  0.16572  $\pm$  0.00096  & 0.16413 $\pm$ 0.00095 & 0.7212 & -0.5332 & 0.8718 & -0.3765 \\
8985$-$10300  &  0.16607  $\pm$  0.00089  & 0.16459 $\pm$ 0.00088 & 0.7134 & -0.5378 & 0.8618 & -0.3737 \\
36000 		  &  0.16305  $\pm$  0.00050  & 0.16305 $\pm$ 0.00050 & 0.4935 & -0.2505 & 0.1831 & -0.0638 \\  
45000 		  &  0.16390  $\pm$  0.00110  & 0.16390 $\pm$ 0.00110 & 0.5344 & -0.5777 & 0.5534  & -0.1991 \\
\enddata
\end{deluxetable*}
\label{tab:tr_spec}
\capstarttrue

\section{Results}
\label{sec:results}

The broadband STIS+\textit{Spitzer} transmission spectrum for WASP-52b corrected for stellar activity and compared to theoretical forward atmospheric models \citep{Goyal17} and past transmission spectrum measurements \citep{Chen17,Louden17} is shown in Figure \ref{fig:tr_spec}. The transmission spectrum shows evidence of Na {\sc i} absorption (5893 \AA) at 2.3$\sigma$ confidence and no observable detection of K {\sc i} absorption. To visualize the amplitude of the spot corrections, we also compare the raw transmission spectrum (before applying the stellar activity correction) and the spot corrected spectrum in Figure \ref{fig:spot_comparison}.

\subsection{Constraints on Na {\sc i} \& K {\sc i}}
\label{sec:constraints}
We inspect the presence and significance of the Na {\sc i} and K {\sc i} features in the WASP-52b transmission spectrum using a grid of spectrophotometric channels ranging in width from 30$-$255 \AA~ in steps of 15 \AA, centered on the Na {\sc i} (5893 \AA) and K {\sc i} (7665 \AA) resonance doublets. The minimum bin size (30 \AA) is defined to include both doublet lines. If a planetary atmospheric signal is present, this binning scheme should demonstrate a gradual decay in the measured transit depth for larger bin sizes. The results of this analysis are shown in Figure \ref{fig:bin_size}.


For the Na {\sc i} feature, we note a gradual decrease in the measured transit depth for bins 30$-$100 \AA~wide. For wider bins, the signal is largely washed out and the transit depth remains unchanged within the uncertainties. This trend is expected when observing a narrow absorption peak and is consistent with the presence of a cloud deck. By measuring the difference in transit depth between the narrowest (30 \AA) bin and the flat transit depth baseline, we detect the core of the Na {\sc i} doublet at 2.3$\sigma$ confidence. In the case of K {\sc i}, we do not see evidence of absorption even in the narrowest spectroscopic channel, suggesting that this feature is either masked by a thick cloud deck or not as abundant in the atmosphere of WASP-52b. For further discussion, see Section \ref{sec:alkali}.


\subsection{Fits to Forward Atmospheric Models}
\label{sec:fits}
We fit the combined STIS+\textit{Spitzer} transmission spectrum to the publicly available grid of forward model transmission spectra \citep{Goyal17} produced using the  $\tt{ATMO}$ 1D radiative-convective equilibrium model (\citealt{[]Amundsen14,Tremblin15,Tremblin16,Drummond16}). The models are generated for the parameters (e.g., mass, radius, gravity, etc.) of WASP-52b. 

The grid\footnote{\url{https://bd-server.astro.ex.ac.uk/exoplanets/WASP-52/}} includes 3,920 model transmission spectra of WASP-52b for five temperatures (1015 K, 1165 K, 1315 K, 1465 K, 1615 K), seven metallicities (0.005,
0.1, 1, 10, 50, 100, 200$\times$ solar), seven C/O ratios (0.15,
0.35, 0.56, 0.70, 0.75, 1.0, and 1.5), four values of the haziness parameter $\alpha_{haze}$ (1, 10, 150, and 1100), and four values of the cloudiness parameter $\alpha_{cloud}$ (0,
0.06, 0.2, and 1). The parameter $\alpha_{haze}$ is a proxy for the haze enhancement factor of small scattering aerosol particles suspended in the atmosphere, where $\alpha_{haze}$=1 indicates no haze and $\alpha_{haze}$=1100 indicates thick hazes. The cloudiness parameter $\alpha_{cloud}$ gives the strength of gray scattering due to H$_{2}$ at 350 nm, with $\alpha_{cloud}$=0 corresponding to no clouds and $\alpha_{cloud}$=1 corresponding to a thick cloud deck. See \citet{Goyal17} and references therein for further details. The transmission spectra are computed assuming isothermal pressure$-$temperature ($P-T$) profiles and condensation without rainout (local condensation). 


We computed the mean model prediction for the wavelength range of each spectroscopic channel (see Table \ref{tab:tr_spec}), and performed a least-squares fitting of the band-averaged model to the spectrum. For the fitting procedure, we allowed the vertical offset in $R_{p}/R_{\star}$ between the spectrum and model to vary while holding all other parameters fixed in order to preserve the model shape. The number of degrees of freedom for each model is $n - m$, where $n$ is the number of data points and $m$ is the number of fitted parameters. Since $n$ = 38 and $m$ = 1, the number of degrees of freedom for each model is constant. From the fits, we computed the $\chi^{2}$ statistic to quantify our model selection. 

Figure \ref{fig:tr_spec} shows the best fit model, representative clear and hazy models, and a flat model compared to the observed transmission spectrum. 
The best fitting model ($\chi^{2}$=39.3) is cloudy ($\alpha_{cloud}$=1.0) and slightly hazy ($\alpha_{haze}$=10) with a 2.3$\sigma$ signature of Na {\sc i} absorption, a temperature of $T$ = 1315 K, solar metallicity ([M/H]=0.0), and slightly super-solar C/O (C/O=0.70). The selected clear model ($\chi^{2}$=49.5) has a lower temperature ($T$ = 1015 K) and no clouds ($\alpha_{cloud}$=0.00) or hazes ($\alpha_{haze}$=1). The representative hazy model ($\chi^{2}$=43.3) is similar to the clear model, but with extreme haziness ($\alpha_{haze}$=1100). The flat model ($\chi^{2}$=52.2) represents a featureless (gray) spectrum.  

\begin{figure*}
\centering
\includegraphics[scale=0.90]{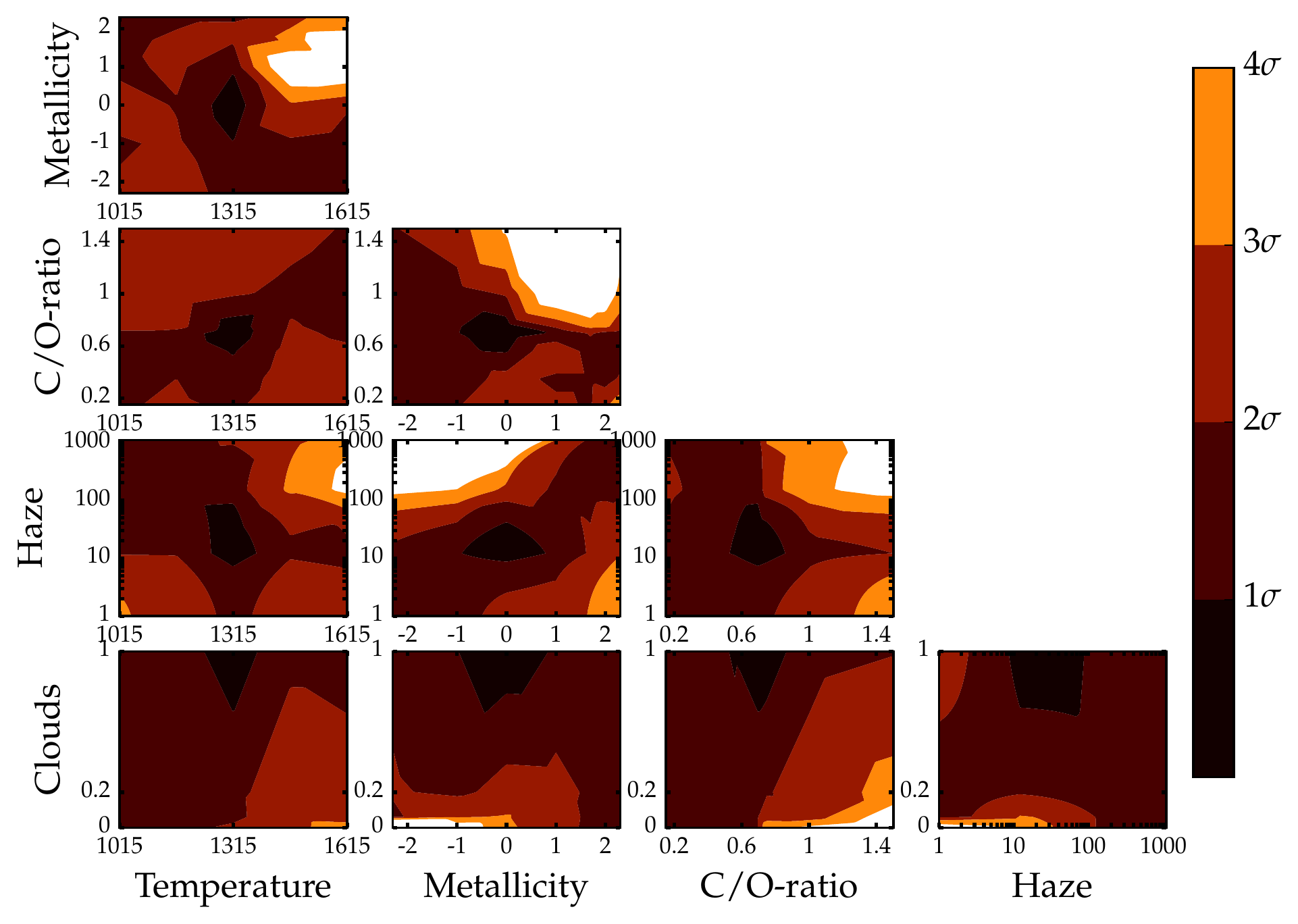}
\centering
\caption{$\chi^{2}$ map for WASP-52b for the no rainout {\tt ATMO} model grid. The cloud, haze, and metallicity axes are log-scaled. The contours show all combinations of the grid parameters, and the colors indicate the confidence intervals corresponding to the colorbar to the right. The white regions correspond to parameter space on the grid that are not feasible given current observations at 4$\sigma$ confidence and can therefore be easily ruled out.}
\label{fig:chi_squared}
\end{figure*}

The $\chi^{2}$ contour plot for the model grid fits is shown in Figure \ref{fig:chi_squared}. The grid provides constraints on the atmospheric and physical parameters of WASP-52b, with the 1$\sigma$ confidence region favoring high cloudiness, slight haziness, solar metallicity, slightly super-solar C/O ratio ($>$0.56) and an equilibrium temperature of 1315 K. 

\capstartfalse

\begin{deluxetable}{ccc}
\tabletypesize{\scriptsize}
\tablewidth{0pt}
\tablecolumns{3}
\tablecaption{System parameters and best fitting model parameters for WASP-52b \&  HAT-P-1b \citep{Nikolov14} \label{tab:comparison}}

\tablehead{\colhead{}  & \colhead{WASP-52b}  & \colhead{HAT-P-1b} }  

\startdata 
Spectral Type   &   K2V  &   G0V \\
Stellar mass~ $M_{\star} ~(M_{\odot})$       	  & 0.87 $\pm$ 0.03 				   &  1.15 $\pm$ 0.05   			   \\
Stellar radius~ $R_{\star} ~(R_{\odot})$  	 	  & 0.79 $\pm$ 0.02	 			  &  1.17 $\pm$ 0.03   			   \\
Surface gravity~ $log(g_{\star})$ ~(cgs)	 	  & 4.58 $\pm$ 0.014    			   &  4.36 $\pm$ 0.01   			   \\
Stellar Temperature~ $T_{eff} ~(K)$     	 	  & 5000 $\pm$ 100	    		   &  5980 $\pm$ 50     			   \\
Metallicity~ $[Fe/H]$          			     	  & 0.03 $\pm$  0.12    			   &  0.13 $\pm$ 0.01   			   \\
Stellar irradiation~ $I$ ~(erg/cm$^{2} \rm s$)	  & 6.5$\pm$0.4 $\times$ 10$^{8}$	   &  7.0$\pm$0.4 $\times$ 10$^{8}$   \\ 
$\rm log(R'_{HK})$  & $-4.4 \pm 0.2$  &  $-4.98 \pm 0.1$ \\
$M_{p}$ ($M_{\rm J}$) 							  & 0.434 $\pm$ 0.024				   &  0.525 $\pm$ 0.019  			   \\
$R_{p}$ ($R_{\rm J}$) 							  & 1.253 $\pm$ 0.027				   &  1.319 $\pm$ 0.019   			   \\
$\rho$ ($\rho_{\rm J}$) 						  & 0.206 $\pm$ 0.009				   &  0.213 $\pm$ 0.010 			   \\
\cutinhead{Best fitting ATMO models}
$T_{\rm eq}$ (K)   & 1315        &  1322   \\
Fe/H 			   & 0.0	     &  1.0    \\
C/O				   & 0.70        &  0.15   \\
$\alpha_{cloud}$    & 1.0	     &  0.20   \\
$\alpha_{haze}$   & 10	         &  10     \\
\enddata
\end{deluxetable}
\label{tab:comparison}
\capstarttrue

\subsection{Comparison with Previous Results}
\label{sec:previous}
In addition to the combined STIS+\textit{Spitzer} transmission spectrum we report here, there are two ground-based optical transmission spectrum measurements for WASP-52b. Most recently, \citet{Louden17} used spectroscopy between 4000 and 8750 \AA~  for two transit observations from the ACAM instrument mounted on the William Herschel Telescope (WHT). \citet{Chen17} observed one transit with the Gran Telescopio Canarias's (GTC) OSIRIS instrument in the 5220$-$9030 \AA~ wavelength range. We show these transmission spectra compared to our STIS+\textit{Spitzer} results in Figure \ref{fig:tr_spec}.  

With the WHT/ACAM observations, \citet{Louden17} modeled spot-crossing events via Gaussian processes, adopted a harmonic analysis of ground-based photometric monitoring, and found varying levels of activity over time with evidence of differential rotation. These results reveal a flat transmission spectrum attributed to an optically thick cloud deck. The GTC/OSIRIS observations indicate a cloudy atmosphere with a $\sim$3$\sigma$ detection of Na {\sc i} and a weaker detection of K {\sc i}. Calculations of the integrated absorption depth for the Na {\sc i} and K {\sc i} signals suggest an inverted temperature structure for the upper atmosphere of WASP-52b \citep{Chen17}. 

Our $R_{p}/R_{\star}$ baseline is consistent within 1$\sigma$ with the ground-based transmission spectrum from WHT/ACAM \citep{Louden17}. The most significant difference is that the WHT spectrum does not show any variation in $R_{p}/R_{\star}$ around 5893 \AA. If there is a weak signal of Na {\sc i} absorption from the planet, the resolution of the spectrum, comprised of equally sized spectrophotometric bins of width 250 \AA, may be washing it out as illustrated in Figure \ref{fig:bin_size}.

The baseline of our spectrum matches less well with the ground-based GTC/OSIRIS transmission spectrum \citep{Chen17}. We note a constant offset in the absolute measured transit depths of our STIS+\textit{Spitzer} spectrum and the GTC measurement, with a difference in the $R_{p}/R_{\star}$ baseline of $\sim$3$\sigma$. The authors attribute their shallower transit depth measurement compared to previous studies (e.g., \citealt{[]Hebrard13,Kirk16,Mancini17}) to the effects of stellar activity. 
We find evidence of a Na {\sc i} signal that is consistent with the GTC detection within 1$\sigma$ but our spectrum shows no evidence of K {\sc i} absorption, contrary to the \citet{Chen17} result. This discrepancy could be due to the different methods used to correct for the effects of stellar activity (see Section \ref{sec:activity_corr} and c.f. \citealt{Chen17}). 


\subsection{Comparison to HAT-P-1b}


We compared WASP-52b to the well-studied inflated hot Jupiter HAT-P-1b (c.f., \citealt{Nikolov14}), since both planets have comparable system parameters and atmospheric properties. HAT-P-1b and WASP-52b have overlapping surface gravity, equilibrium temperature, mass, radius, and stellar irradiation, and the transmission spectra of both planets are marginally flat with evidence of Na {\sc i} absorption but no observable K {\sc i} absorption. Table \ref{tab:comparison} compares the stellar and planetary parameters for WASP-52 (\citealt{Hebrard13}; Table \ref{tab:params}) and HAT-P-1 (\citealt{Nikolov14}).

HAT-P-1b has a precise transmission spectrum measurement from \textit{HST}/STIS \citep{Nikolov14}, which we use to compare the atmospheric properties of both planets. For this comparison, we reconstructed the \textit{HST}/STIS transmission spectrum for WASP-52b using the same binning scheme of \citet{Nikolov14} for HAT-P-1b and fit the light curves for these bins based on the methods outlined in Section \ref{sec:analysis}. Figure \ref{fig:trspec_comp} shows the \textit{HST}/STIS transmission spectra of both planets with identical binning. Based on this comparison, the spectra of both planets are identical within the uncertainties with an average 1$\sigma$ difference.  

As reported in \citet{Nikolov14}, the best fit model for HAT-P-1b is a hazy spectrum with Na {\sc i} absorption and an extra optical absorber to account for the observed absorption enhancement at wavelengths longer than $\sim$0.85 $\mu$m. To directly compare this interpretation to the WASP-52b results reported here, we fit the HAT-P-1b spectrum \citep{Nikolov14} to the open source {\tt ATMO} grid of forward models generated for the parameters of HAT-P-1b (see Appendix \ref{sec.hatp1} and Figure \ref{fig:H1_fits} for details). 
The best fit model parameters for HAT-P-1b and WASP-52b are shown in Table \ref{tab:comparison}. 

Since WASP-52b and HAT-P-1b have similar optical transmission spectra ($\sim$0.29$-$1 $\mu$m), we compare the atmospheric properties of these planets in the near-infrared to ascertain if these planets would be good candidates for comparative atmospheric analyses with \textit{JWST}. Figure \ref{fig:H1_comparison} shows the best fit models for both planets from 0.29$-$5 $\mu$m. Beyond 1 $\mu$m, we find that the best fitting models for HAT-P-1b and WASP-52b do not agree with each other as they do in the optical. For further discussion regarding potential reasons for this near-infrared discrepancy, see Section \ref{sec:contextualizing}. 

\section{Discussion}
\label{sec:discussion}

\subsection{Interpreting Alkali Detections in the Presence of Clouds}
\label{sec:alkali}

As reported in Section \ref{sec:constraints}, we find hints of Na {\sc i} absorption but no evidence of K {\sc i} in the transmission spectrum of WASP-52b. A similar trend has been observed for HD 189733b \citep{Pont13}, HAT-P-1b \citep{Nikolov14}, and WASP-17b \citep{Sing16}. The reverse trend (i.e., the presence of a K {\sc i} signal but no Na {\sc i}) has been observed in several planets, including WASP-31b \citep{Sing15} and HAT-P-12b \citep{Sing16}. We note, however, that the majority of current sodium and potassium detections in exoplanet atmospheres are low significance and a non-detection of K {\sc i} can only be interpreted as an upper limit to the abundance of that element in the atmospheric layers probed by our observations, considering their uncertainties. Based on the available data and their uncertainties, we estimate an abundance ratio of ln[Na/K] =  $8.32^{+ 6.51}_{- 6.03}$. This value is consistent with solar abundances but has large uncertainties and is not well-constrained. 

These results are consistent with our forward model analysis, which favors the presence of clouds and slight hazes that mute spectroscopic features in the planet's atmosphere. The best fitting models give temperatures ranging from $T\sim$ 1000$-$1300 K. Compared to the planet's equilibrium temperature ($T_{\rm eq}$ = 1315 K; \citealt{Hebrard13}), these temperature estimates may be driven by the gradient of the Rayleigh scattering slope. Significant sodium condensation in the form of Na$_{2}$S is expected at lower temperatures ($\sim$1000 K) and potassium condensation in the form of KCl at even lower temperatures ($\sim$600 K) \citep{Marley13}. Tenuous Na$_{2}$S and KCl clouds could form at shallow pressures in WASP-52b's atmosphere, since the equilibrium temperature does not represent the full range of temperatures in a planet's atmosphere. The cloud deck in WASP-52b's atmosphere could therefore be comprised of species other than sodium or potassium compounds, such as silicates. 

\begin{figure*}
\centering
\includegraphics[scale=0.70]{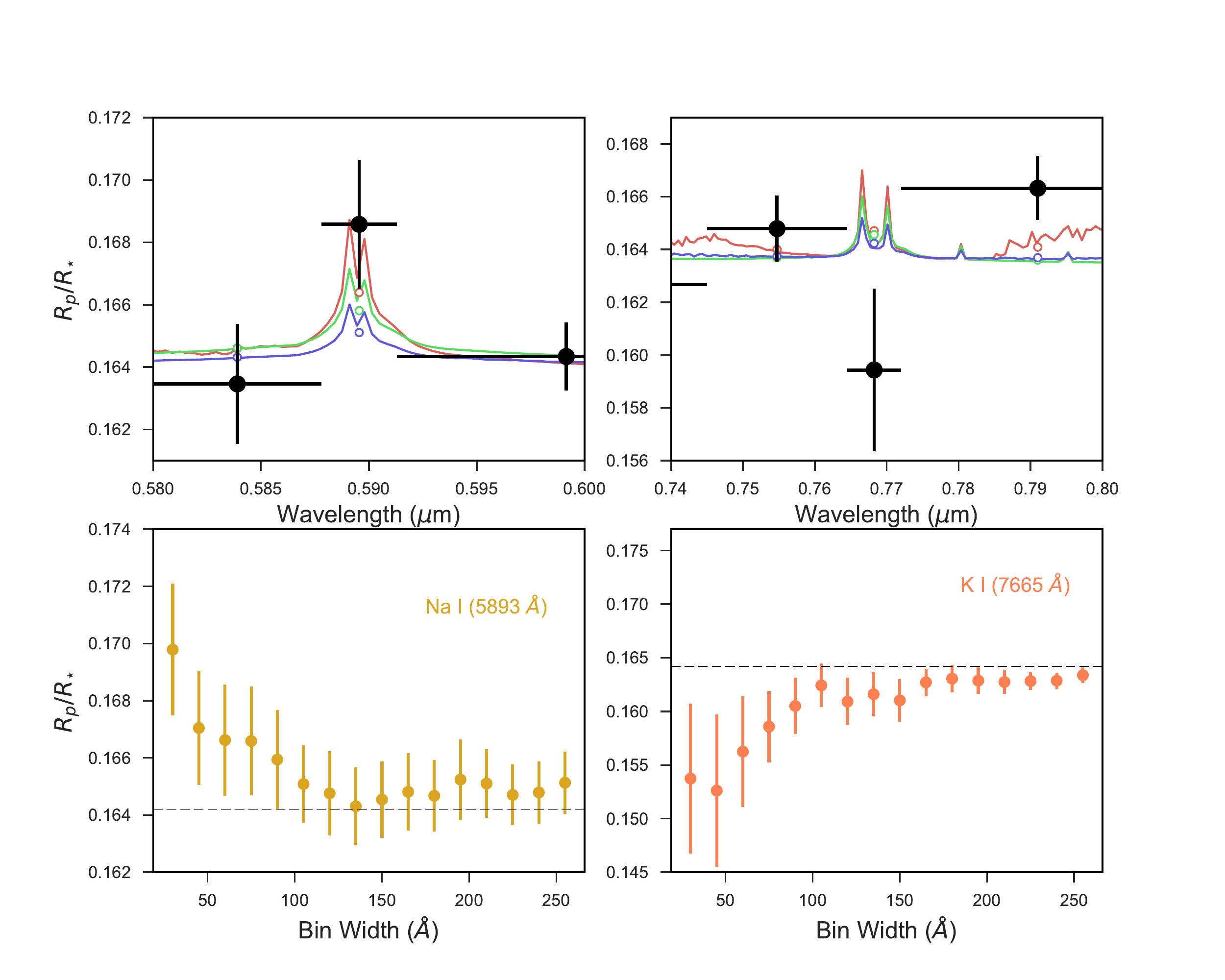}
\centering
\caption{\textit{Top panel}: WASP-52b transmission spectrum (black points) zoomed in to the central wavelength of Na {\sc i} at 5893 \AA~ (left) and K {\sc i} at 7665 \AA~ (right) with the best fit (blue), hazy (green) and clear (red) models overplotted. \textit{Bottom panel}: Absorption depth of Na {\sc i} at 5893 \AA~ (left) and K {\sc i} at 7665 \AA~ (right) for spectroscopic channels ranging in size from 30-255 \AA~. The average $R_{p}/R_{\star}$ baseline of the transmission spectrum (dashed black line) is shown for reference.}
\label{fig:bin_size}
\end{figure*}

Regardless of the composition of these clouds, they are likely masking the K {\sc i} feature and the wings of the Na {\sc i} resonance core. We do not resolve the broad wings of the Na {\sc i} line (Figure \ref{fig:bin_size}), which may suggest the presence of an extra absorber or scatterer in the atmosphere that is obscuring or masking the atmospheric Na {\sc i} and K {\sc i} absorption features (\citealt{Seager00,Nikolov14}). 

If a K {\sc i} signal is truly lacking in the transmission spectrum of WASP-52b, an alternative explanation could be attributed to an underabundance of this element in the planetary atmosphere. If the cloud deck is comprised of sodium or potassium compounds, however, the gas phase abundances of these species would not reflect primordial abundances. Since K {\sc i} is a weaker spectroscopic feature, it may be present in the planet's atmosphere but not detectable with the precision of the STIS data. Higher resolution, higher precision observations are necessary to confirm this idea.


\subsection{Contextualizing WASP-52b}
\label{sec:contextualizing}

Although the atmospheres of planets studied thus far appear diverse \citep{Sing16}, we have not yet been able to identify any clear correlations between planetary atmospheric properties and other system parameters. Comparing the transmission spectra of planets with similar system parameters may therefore prove insightful in searching for common atmospheric characteristics. WASP-52b is a good target for such comparisons, since the well-studied inflated hot Jupiter HAT-P-1b has comparable system parameters and atmospheric properties. 

We compare the observed transmission spectrum of WASP-52b presented here to that of the well-studied inflated hot Jupiter HAT-P-1b. These two planets have similar system parameters (see Table \ref{tab:comparison}), although WASP-52 is $\sim$0.9 times more active that HAT-P-1b \citep{Nikolov14}. Additionally, the optical transmission spectra of both planets are marginally flat with evidence of Na {\sc i} but no observable evidence of K {\sc i}. We compare their transmission spectra with identical binning schemes in Figure \ref{fig:trspec_comp}, and find that the spectra of both planets in the optical ($\sim$0.29$-$1 $\mu$m) are identical within the uncertainties with an average 1$\sigma$ difference. 

Extending this comparison to near-infrared wavelengths, however, reveals that their transmission spectra differ considerably beyond 1 $\mu$m. The best fit {\tt ATMO} model for WASP-52b shows muted H$_{2}$O and CH$_{4}$ spectral features compared to the best fitting HAT-P-1b model, which could indicate that WASP-52b has a higher aerosol layer. Near-infrared \textit{HST}/WFC3 observations of HAT-P-1b reveal H$_{2}$O absorption at $>$5$\sigma$ confidence \citep{Wakeford13}, and the deep H$_{2}$O feature shown in the best fit {\tt ATMO} model matches these data. The best fit atmospheric model for WASP-52b shows a weaker (but still observable) H$_{2}$O feature at 1.4 $\mu$m, and evidence of H$_{2}$O absorption at 1.4 $\mu$m for WASP-52b has recently been shown in \citet{Bruno18}. The activity levels of the host stars may also contribute to the divergence of the transmission spectra for these two planets at near-infrared wavelengths, although observations of the stellar UV fluxes are necessary to confirm this hypothesis. Comparative near-infrared observations with \textit{JWST} can confirm the atmospheric similarities of these planets at shallower atmospheric layers compared to those probed by STIS.

\section{Summary}
\label{sec:conclusions}

\begin{figure*}
\centering
\includegraphics[scale=0.60]{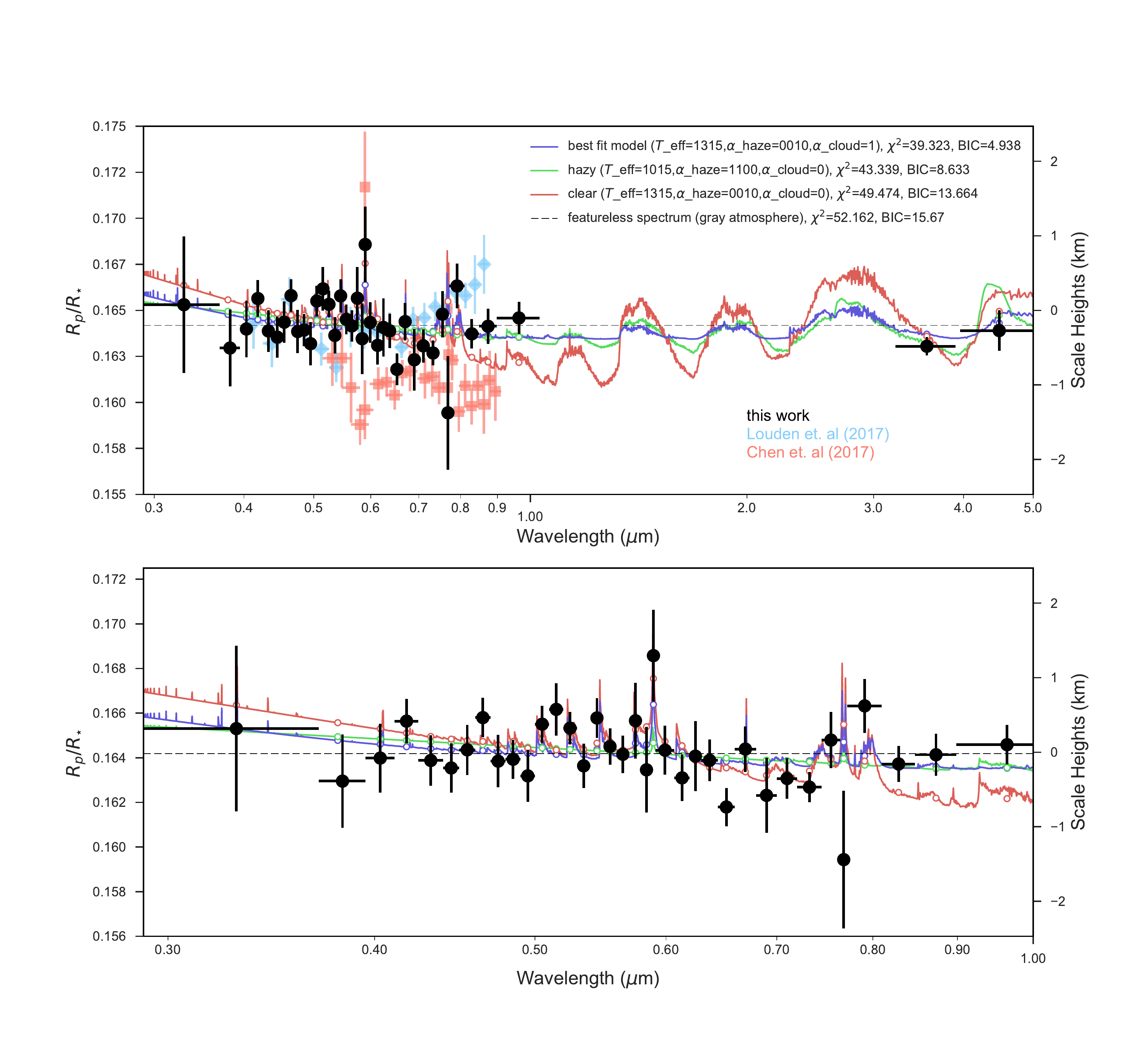} 
\centering
\caption{\textit{Top panel:} Stellar activity corrected transmission spectrum for WASP-52b from \textit{HST}/STIS and \textit{Spitzer}/IRAC (black circles). Ground-based optical transmission spectrum measurements from \citealt{Louden17} (blue diamonds) and \citealt{Chen17} (salmon rectangles) are included for comparison. We show a subset of the best-fit theoretical atmospheric models (lines), and find that the observed transmission spectrum is consistent with evidence of Na {\sc i} at 2.3$\sigma$ confidence and a cloudy atmosphere with no TiO. Models are smoothed by a constant. The average $R_{p}/R_{\star}$ baseline of the transmission spectrum (dashed black line) is shown for reference. \textit{Bottom panel:} Same as above, but zoomed in to the STIS wavelength range ($\sim$0.29-1.0 $\mu$m).}
\label{fig:tr_spec}
\end{figure*}

\begin{figure*}
\centering
\includegraphics[scale=0.60]{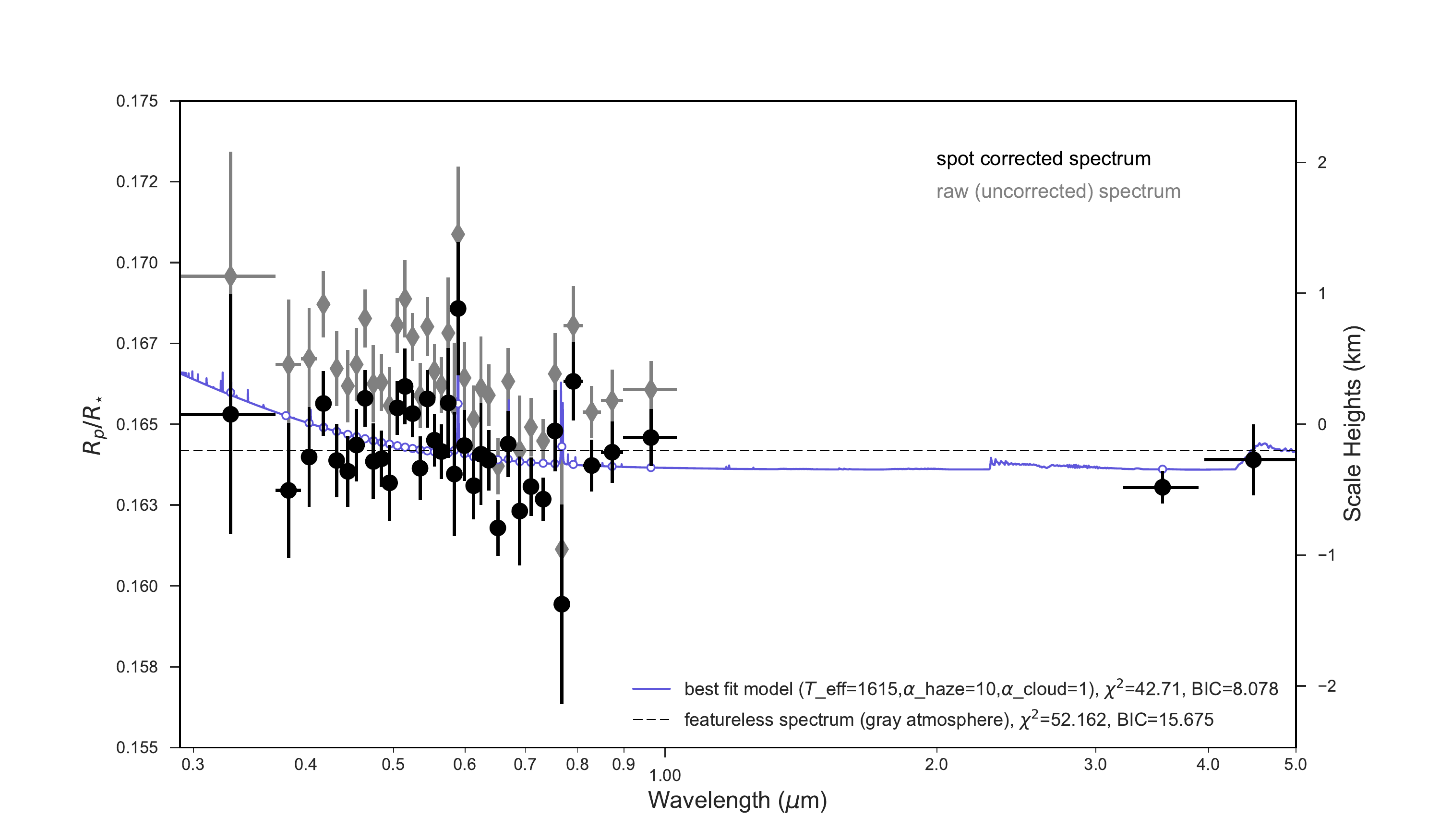}
\centering
\caption{Comparison of the raw (gray diamonds) and spot corrected (black circles) transmission spectra for WASP-52b. The best fitting (solid blue line) and gray atmosphere (dashed black line) models from Figure \ref{fig:tr_spec} are shown for reference.}
\label{fig:spot_comparison}
\end{figure*}

\begin{figure*}
\centering
\includegraphics[scale=0.75]{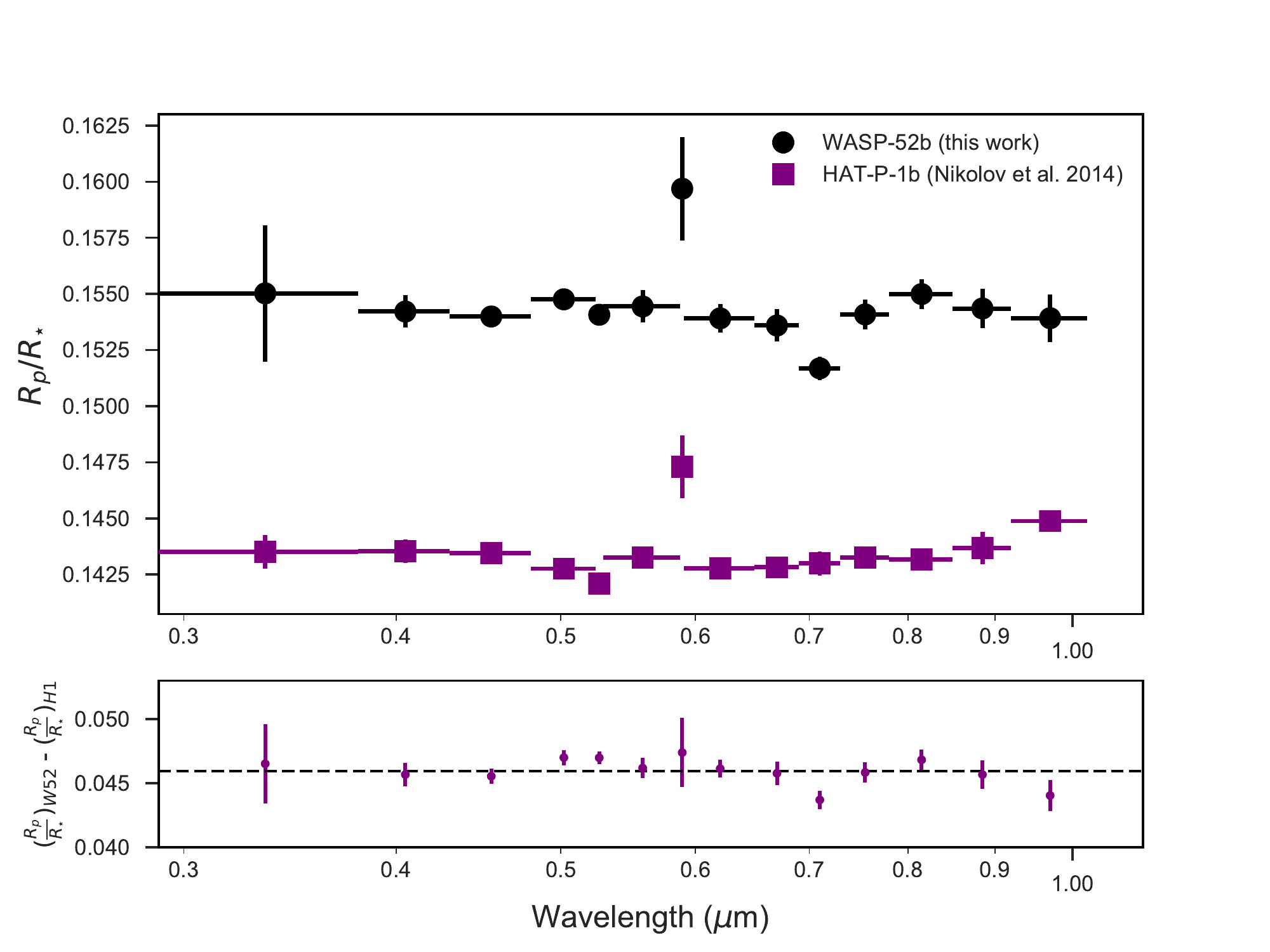}
\centering
\caption{\textit{Top panel}: Comparison of the observed STIS transmission spectra of WASP-52b (black circles) and HAT-P-1b (purple squares; \citealt{Nikolov14}) for identical wavelength bins. The HAT-P-1b spectrum is offset by an arbitrary constant such that the first spectroscopic bin is anchored to the first $R_{p}/R_{\star}$ value of the WASP-52b spectrum. \textit{Bottom panel}: Difference between the observed WASP-52b spectrum and the offset HAT-P-1b spectrum. The spectra of both planets are identical within the uncertainties (except for one channel at $\sim$0.7 $\mu$m), with an average 1$\sigma$ difference. The horizontal black dashed line indicates the baseline offset between the two spectra.}
\label{fig:trspec_comp}
\end{figure*}

\begin{figure*}
\centering
\includegraphics[scale=0.80]{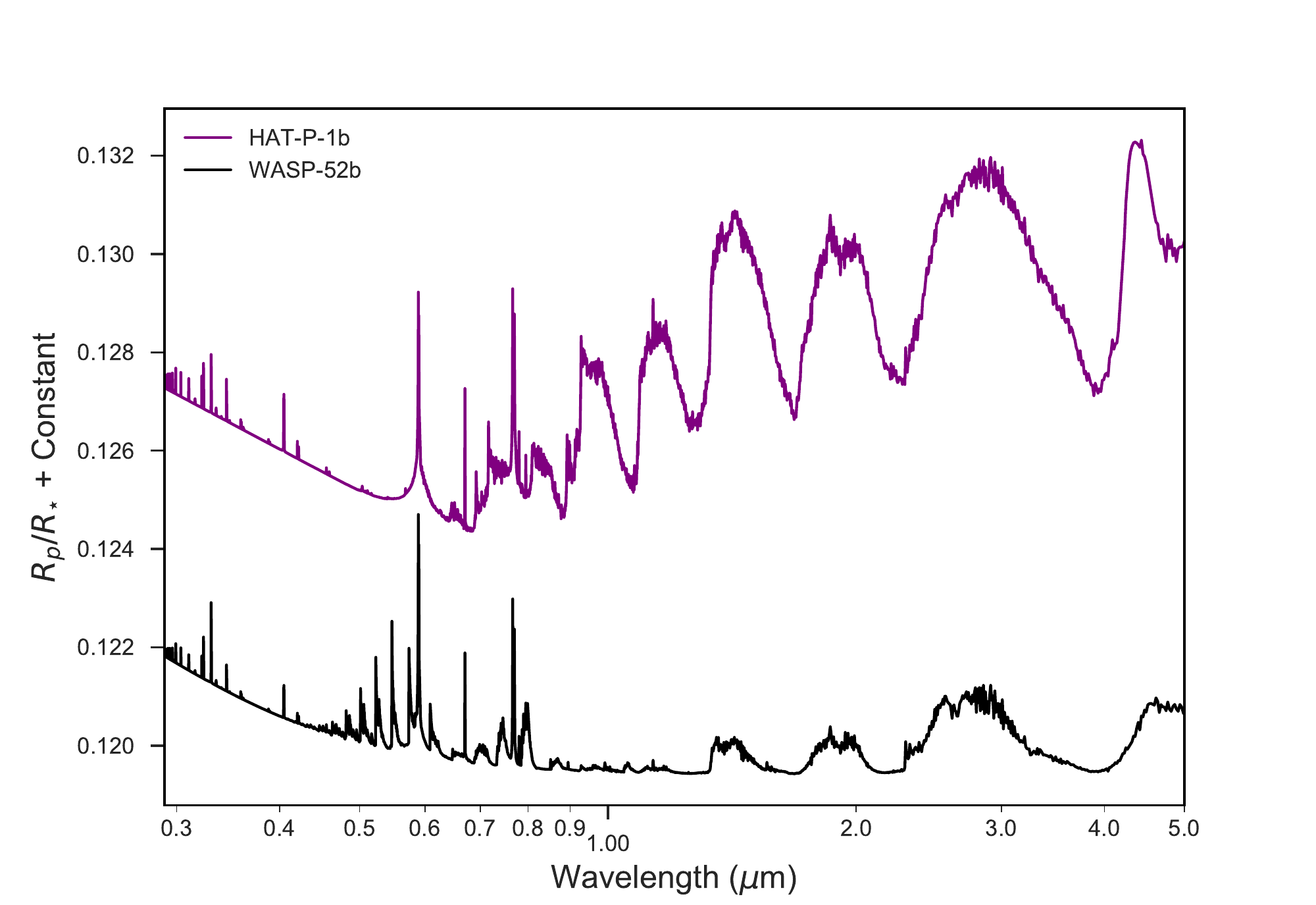}
\centering
\caption{Comparison of the best fit models for HAT-P-1b (purple) and WASP-52b (black) from $\sim$0.29 to 5 $\mu$m. The spectra are identical within the uncertainties at optical wavelengths but differ in the near-infrared.}
\label{fig:H1_comparison}
\end{figure*}

Our key results are summarized as follows:

\begin{itemize}
\item We present an optical to near-infrared transmission spectrum of WASP-52b (measured to a median precision of 90 ppm) from $\sim$0.29 $-$ 5.0 $\mu$m using transit observations from \textit{HST}/STIS and \textit{Spitzer}/IRAC. We correct for the effects of stellar activity and fit the observed transmission spectrum to a grid of forward atmospheric models. 

\item Based on these fits (Figure \ref{fig:tr_spec}), we find that our transmission spectrum measurement best matches a moderately cloudy atmospheric model with an equilibrium temperature of 1315 K, a thick cloud deck ($\alpha_{cloud}$ = 1.00), a slight Rayleigh scattering slope in the blue ($\alpha_{haze}$ = 10), and hints of a 2.3$\sigma$ Na {\sc i} signal at 5893 \AA. Within the precision of our observations, we do not detect K {\sc i} absorption.

\item We compare the observed transmission spectra of HAT-P-1b and WASP-52b, two planetary systems with similar stellar and planetary parameters (Table \ref{tab:comparison}). 
By constructing optical \textit{HST}/STIS transmission spectra with similar binning schemes (Figure \ref{fig:trspec_comp}), we find that the spectra of these two planets are identical within the uncertainties at optical wavelengths but differ in the near-infrared (Figure \ref{fig:H1_comparison}) based on our best fit models.

\item The difference in the transmission spectra of WASP-52b and HAT-P-1b from $\sim$1.0 $-$ 5.0 $\mu$m may be caused by the presence of an extra optical absorber in the atmosphere of HAT-P-1b \citep{Nikolov14} or uncertainties in the best fitting models, which are isothermal and therefore cannot accurately capture cloud formation. 

\item Comparative atmospheric observations with \textit{JWST} for WASP-52b and HAT-P-1b will be key to understanding planets with similar system parameters and overlapping atmospheric properties. 


\end{itemize}

In a forthcoming paper, we aim to combine the STIS+\textit{Spitzer} transmission spectrum presented here with existing near-infrared \textit{HST}/WFC3 observations \citep{Bruno18}. Using the full optical to near-infrared transmission spectrum, we will retrieve the planet's atmospheric properties  (Bruno, Alam, et al., in prep.) to better constrain the atmospheric structure and chemical composition of this inflated hot Jupiter as well as precisely estimate the Na {\sc i} and K {\sc i} abundances in the planet's atmosphere. Such an analysis will indicate if comparative planetology and comparative atmospheric studies of WASP-52b with future \textit{JWST} observations will prove insightful.

\acknowledgments
The authors thank the anonymous referee for helpful comments that greatly improved this manuscript. This paper makes use of observations
from the NASA/ESA \textit{Hubble Space Telescope}, obtained at the Space Telescope Science Institute, which is operated by the Association of Universities for Research in Astronomy, Inc., under NASA contract NAS 5-26555. These observations are associated with program GO 14767. The research leading to these results has received funding from the European Research Council under the European Union's Seventh Framework Programme (FP7/2007-2013)/ERC grant agreement number 336792. We are thankful to Raphaelle Haywood, James Kirk, Chani Nava, and Ian Weaver for useful discussions. 

MKA acknowledges support by the National Science Foundation through a Graduate Research Fellowship. GWH and MHW acknowledge support from Tennessee State University and the State of Tennessee through its Centers of Excellence program. JSF acknowledges funding by the Spanish MINECO grant AYA2016- 79425-C3-2-P. JKB acknowledges support from the Royal Astronomical Society. VB and DE have received funding from the European Research Council (ERC) under the European Unions Horizon 2020 research and innovation program (project Four Aces; grant agreement no. 724427).

\bibliographystyle{yahapj}
\bibliography{refs}
	
\clearpage



\newpage

\newpage

\newpage

\newpage

\newpage

\newpage

\newpage

\newpage

\newpage

\newpage

\newpage

\clearpage
\appendix\section{Stellar Activity Correction}

\subsection{Kernel for Gaussian Process Regression Model}\label{sec.kernel}

We used a Gaussian process (GP) regression to model the ground-based stellar activity monitoring data (\citealt{Pont13,Haywood15,Aigrain16,Angus17}). In our GP analysis, we used a three component kernel of the form \textbf{K} = $k_{1}$ + $k_{2}$ + $k_{3}$. The $k_{1}$ term models the flexible (quasi-) periodicity of the ground-based activity monitoring data. It is a squared exponential kernel multiplied by an exponential sine squared kernel ($k_{1}$ = $ A^{2} [k(r^{2}) \times k(\textbf{x}_{i},\textbf{x}_{j})]$), where 

\begin{equation}
k(r^{2}) = e^{ r^{2}/2 } 
\label{eq:B1}
\end{equation}
represents the squared exponential kernel for periodic variations in the time series data parametrized by $r$.  The exponential sine squared kernel is given by 
\begin{equation}
k(\textbf{x}_{i},\textbf{x}_{j}) = \textit{exp} \Big( -\Gamma sin^2 \Big[ \frac{\pi}{P} |{x}_{i} - {x}_{j}| \Big] \Big)
\label{eq:B2}
\end{equation}
where $\Gamma$ represents the scale of the correlations and $P$ is the period of the oscillations. 

The second term, $k_{2}$, represents the irregularities in amplitude and period. It is a rational quadratic kernel of the form: 
\begin{equation}
k_{2}(r^{2}) = A^{2} \Big( 1 - \frac{r^{2}}{2 \alpha} \Big)
\label{eq:B2}
\end{equation}
for amplitude $A$, Gamma distribution parameter $\alpha$, and length scale $r$. 

The $k_{3}$ term incorporates stellar noise in the GP model and is a squared exponential kernel (of the form shown in Equation \ref{eq:B1}) added to a white kernel of the form $k_{3}(\textbf{x}_{i},\textbf{x}_{j}) = c \delta_{ij}$ for constant $c$ and diagonal value $\delta_{ij}$. 


\subsection{Deriving the Wavelength-Dependent Flux Correction}\label{sec.corr}

As described in Section \ref{sec:activity_corr}, we account for the effects of stellar activity and unocculted starspots on the transmission spectrum by using ground-based photometry and by deriving the wavelength-dependent flux correction $\Delta f(\lambda,T)$ (\citealt{Sing11}; see also Section \ref{sec:corr}, Equation \ref{eqn:flux_corr}). This correction depends on the parameter $k$, which provides an assumption for the non-spotted flux on the stellar surface \citep{Aigrain12}. We fixed $k$ to unity based on \citet{Aigrain12}, which found this value appropriate to use for active stars. Physically, $k$=1 corresponds to a spot contribution that the viewer never sees (or always sees) that is about the same as the contribution of spots that come into and out of view.

\begin{figure*}
\centering
\includegraphics[scale=0.70]{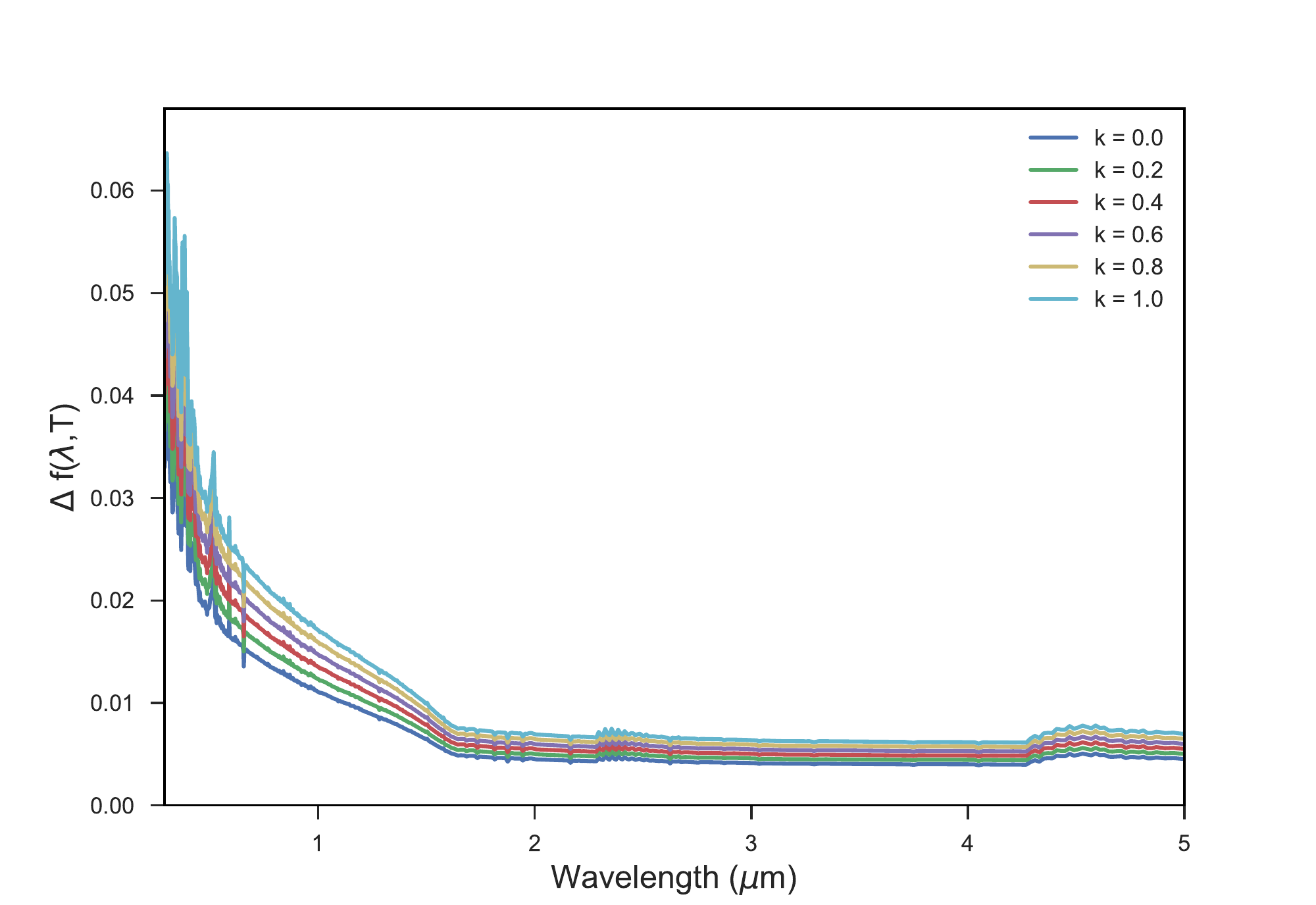}
\centering
\caption{Wavelength-dependent flux correction for a fixed spot temperature (4750 K) and different values of the parameter $k$, a proxy for the non-spotted fraction of the stellar surface. For the activity correction described in Section \ref{sec:corr}, we assume $k$=1 (i.e., the spot contribution that the viewer never sees (or always sees) is about the same as the contribution of spots that come into and out of view).}
\label{fig:k_test}
\end{figure*}

The value of this parameter is based on several assumptions; however, it is, in actuality, very ill-constrained and we warn the reader of the difficulty in selecting a value of $k$. To explore the potential effect of the choice of $k$ on the final transmission spectrum, we tested different values of this parameter ranging from $k$=0$-$1 in steps of 0.2 (Figure \ref{fig:k_test}). We find that larger values of $k$ correspond (linearly) to a higher derived flux correction $\Delta f(\lambda, T)$. When applying the activity correction to the binned light curves, we find that the light curves are re-scaled to a higher $R_{p}/R_{\star}$ baseline if a larger correction is applied. Thus, the parameter $k$ is not wavelength-dependent and therefore does not affect the shape of the slope of the transmission spectrum. Rather, varying the value of $k$ simply shifts the $R_{p}/R_{\star}$ baseline of the transmission spectrum vertically.

\section{STIS Light Curves}

\subsection{STIS White Light Curve Systematics Models}\label{sec.sys}

\setcounter{table}{0} \renewcommand{\thetable}{B\arabic{table}}

As described in Section \ref{sec:stis_wlc}, we detrended instrument systematic effects in the light curves by fitting a fourth-order polynomial to the flux dependence on \textit{HST} orbital phase. Each model represents a unique linear combination of the detrending variables: orbital phase ($\phi_{\rm{t}}$), drift of the spectra on the detector ($x$ and $y$), the shift ($\omega$) of each stellar spectrum cross-correlated with the first spectrum of the time series, and time $t$. The variables $x$ and $y$ are the trace slope and an offset in the cross dispersion direction, respectively. The parameter $\omega$ is measured by cross-correlating a reference spectrum with the remaining spectra and is measured prior to re-sampling the spectra.

For both the G430L and G750L STIS observations, we used the 25 systematics models listed in Table \ref{tab:systematics} to detrend the light curves. After performing separate fits for each model, we marginalized over the entire set of systematics models assuming equally weighted priors to select which systematics model to use. Table \ref{tab:systematic_models} summarizes the selection of systematics models based on the STIS white light curves. We selected the model for detrending based on the lowest Aikake Information Criterion (AIC) value. 

\subsection{STIS Spectrophotometric Light Curves}

The broadband STIS+\textit{Spitzer} transmission spectrum for WASP-52b reported in Table \ref{tab:tr_spec} gives the weighted mean of the two STIS G430L observations (visits 52 and 53). In Tables \ref{tab:v52} and \ref{tab:v53} below, we report the spectrophotometric light curves for each G430L visit. To produce the raw transmission spectrum from the spot corrected spectrum, the reader may simply reverse the correction given in Table \ref{tab:fnorm} on each transit observation separately.


\section{HAT-P-1$\MakeLowercase{{\rm b}}$ Forward Model Fits}\label{sec.hatp1}
\renewcommand\thefigure{\thesection.\arabic{figure}}  
\setcounter{figure}{0} 

\begin{figure*}
\centering
\includegraphics[scale=0.60]{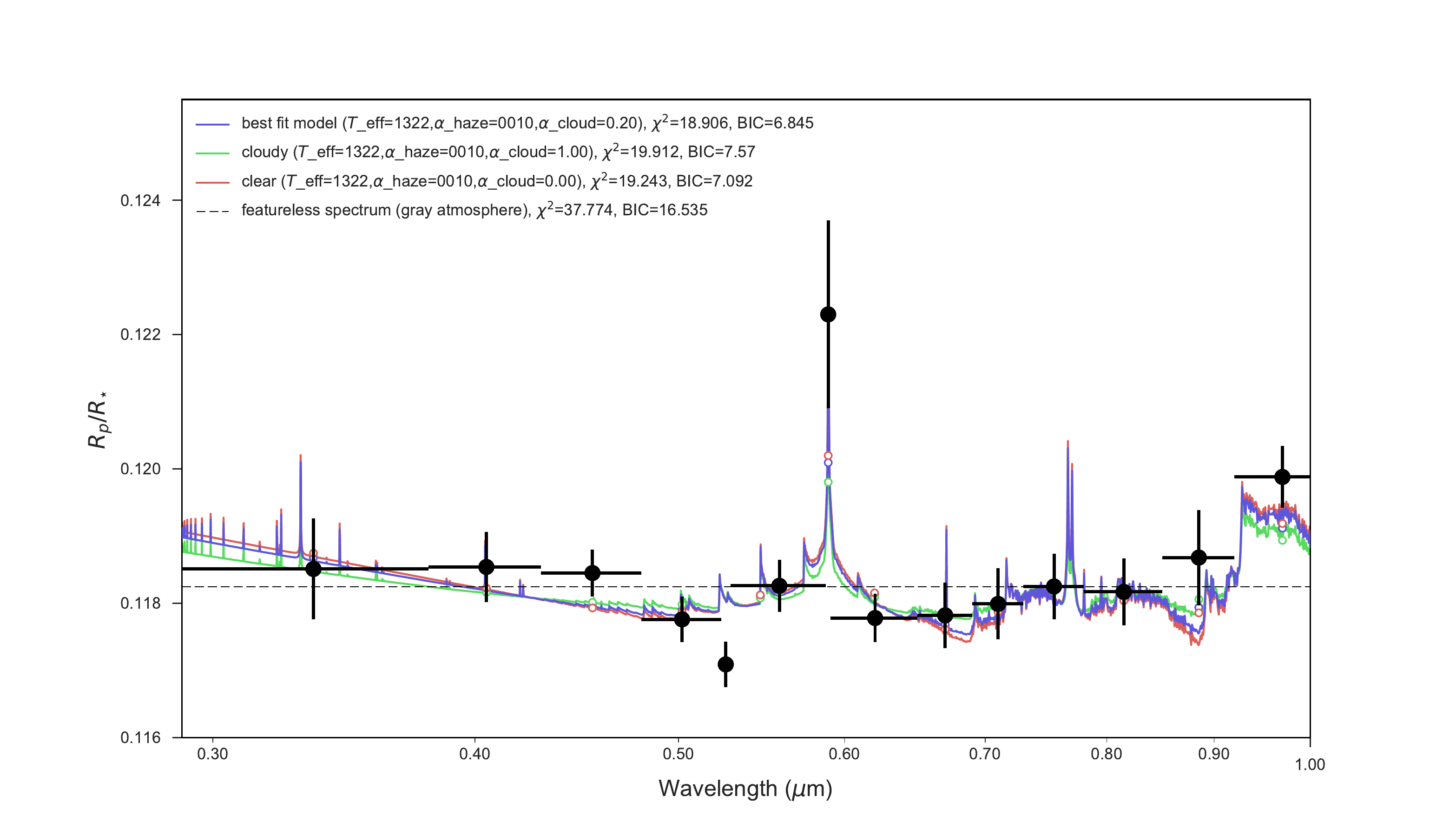}
\centering
\caption{\textit{HST}/STIS transmission spectrum of HAT-P-1b (black points; \citealt{Nikolov14}) compared to a subset of the best-fit theoretical atmospheric models (lines) for the no rainout {\tt ATMO} grid. For reference, the dashed black line shows the average $R_{p}/R_{\star}$ baseline of the transmission spectrum.}
\label{fig:H1_fits}
\end{figure*}

Figure \ref{fig:H1_fits} shows the \textit{HST}/STIS transmission spectrum of the well-studied inflated hot Jupiter HAT-P-1b from \citet{Nikolov14} compared to the best fitting theoretical forward model from the ATMO grid generated for the parameters of HAT-P-1b\footnote{\url{https://bd-server.astro.ex.ac.uk/exoplanets/HAT-P-01/}} in addition to representative clear and cloudy models. We performed these fits using the procedure described in Section \ref{sec:fits}. We find that the best fit model has $T$ = 1322 K with a strong Na {\sc i} signal and is slightly cloudy ($\alpha_{cloud}$=0.20) and hazy ($\alpha_{haze}$=10). 

\newpage
\capstartfalse

\startlongtable
\begin{deluxetable}{cl}
\tabletypesize{\scriptsize}
\tablewidth{0pt}
\tablecolumns{2}
\tablecaption{White Light Curve Systematics Models  
\label{tab:systematics}}

\tablehead{\colhead{ }  & \colhead{Model}  }
\startdata 
\cutinhead{G430L models}
   1 & \multicolumn{1}{l}{$\phi_{t}+\phi^{2}_{t}+\phi^{3}_{t}+\phi^{4}_{t}+t$}   \\
   2 & \multicolumn{1}{l}{$\phi_{t}+\phi^{2}_{t}+\phi^{3}_{t}+\phi^{4}_{t}+t+\omega+x^2$}   \\
   3 & \multicolumn{1}{l}{$\phi_{t}+\phi^{2}_{t}+\phi^{3}_{t}+\phi^{4}_{t}+t+x+y^2$}  \\
   4 & \multicolumn{1}{l}{$\phi_{t}+\phi^{2}_{t}+\phi^{3}_{t}+\phi^{4}_{t}+t+x^2+y$ } \\
   5 & \multicolumn{1}{l}{$\phi_{t}+\phi^{2}_{t}+\phi^{3}_{t}+\phi^{4}_{t}+t+\omega$}  \\
   6 & \multicolumn{1}{l}{$\phi_{t}+\phi^{2}_{t}+\phi^{3}_{t}+\phi^{4}_{t}+t+x$}  \\
   7 & \multicolumn{1}{l}{$\phi_{t}+\phi^{2}_{t}+\phi^{3}_{t}+\phi^{4}_{t}+t+y$}  \\
   8 & \multicolumn{1}{l}{$\phi_{t}+\phi^{2}_{t}+\phi^{3}_{t}+\phi^{4}_{t}+t+\omega+\omega^3+x$}  \\
   9 & \multicolumn{1}{l}{$\phi_{t}+\phi^{2}_{t}+\phi^{3}_{t}+\phi^{4}_{t}+t+\omega+y$}  \\
  10 & \multicolumn{1}{l}{$\phi_{t}+\phi^{2}_{t}+\phi^{3}_{t}+\phi^{4}_{t}+t+\omega+x+y$}  \\
  11 & \multicolumn{1}{l}{$\phi_{t}+\phi^{2}_{t}+\phi^{3}_{t}+\phi^{4}_{t}+t+\omega+\omega^2$}  \\
  12 & \multicolumn{1}{l}{$\phi_{t}+\phi^{2}_{t}+\phi^{3}_{t}+\phi^{4}_{t}+t+\omega+\omega^2+\omega^3$}  \\
  13 & \multicolumn{1}{l}{$\phi_{t}+\phi^{2}_{t}+\phi^{3}_{t}+\phi^{4}_{t}+t+x+x^2+y$}  \\
  14 & \multicolumn{1}{l}{$\phi_{t}+\phi^{2}_{t}+\phi^{3}_{t}+\phi^{4}_{t}+t+\omega+x+x^2+x^3$}  \\
  15 & \multicolumn{1}{l}{$\phi_{t}+\phi^{2}_{t}+\phi^{3}_{t}+\phi^{4}_{t}+t+\omega+y+y^2$}  \\
  16 & \multicolumn{1}{l}{$\phi_{t}+\phi^{2}_{t}+\phi^{3}_{t}+\phi^{4}_{t}+t+\omega+y+y^2+y^3$}  \\
  17 & \multicolumn{1}{l}{$\phi_{t}+\phi^{2}_{t}+\phi^{3}_{t}+\phi^{4}_{t}+t+\omega+\omega^2+\omega^3+x$}  \\
  18 & \multicolumn{1}{l}{$\phi_{t}+\phi^{2}_{t}+\phi^{3}_{t}+\phi^{4}_{t}+t+x+x^2$}  \\
  19 & \multicolumn{1}{l}{$\phi_{t}+\phi^{2}_{t}+\phi^{3}_{t}+\phi^{4}_{t}+t+x+x^2+x^3$}  \\
  20 & \multicolumn{1}{l}{$\phi_{t}+\phi^{2}_{t}+\phi^{3}_{t}+\phi^{4}_{t}+t+y+y^2$}  \\
  21 & \multicolumn{1}{l}{$\phi_{t}+\phi^{2}_{t}+\phi^{3}_{t}+\phi^{4}_{t}+t+y+y^2+y^3$}  \\
  22 & \multicolumn{1}{l}{$\phi_{t}+\phi^{2}_{t}+\phi^{3}_{t}+\phi^{4}_{t}+t+\omega+\omega^2+x$}  \\
  23 & \multicolumn{1}{l}{$\phi_{t}+\phi^{2}_{t}+\phi^{3}_{t}+\phi^{4}_{t}+t+\omega+\omega^2+x+y$}  \\
  24 & \multicolumn{1}{l}{$\phi_{t}+\phi^{2}_{t}+\phi^{3}_{t}+\phi^{4}_{t}+t+\omega+\omega^2+x+x^2+y+y^2$}  \\
  25 & \multicolumn{1}{l}{$\phi_{t}+\phi^{2}_{t}+\phi^{3}_{t}+\phi^{4}_{t}+t+\omega^2+\omega^3+x+x^2+x^3+y+y^2+y^3$}  \\
\cutinhead{G750L models}
   1 & \multicolumn{1}{l}{$\phi_{t}+\phi^{2}_{t}+\phi^{3}_{t}+\phi^{4}_{t}+t$}  \\
   2 & \multicolumn{1}{l}{$\phi_{t}+\phi^{2}_{t}+\phi^{3}_{t}+\phi^{4}_{t}+t+\omega+x^2$}  \\
   3 & \multicolumn{1}{l}{$\phi_{t}+\phi^{2}_{t}+\phi^{3}_{t}+\phi^{4}_{t}+t+x+y^2$}  \\
   4 & \multicolumn{1}{l}{$\phi_{t}+\phi^{2}_{t}+\phi^{3}_{t}+\phi^{4}_{t}+t+x^2+y$}  \\
   5 & \multicolumn{1}{l}{$\phi_{t}+\phi^{2}_{t}+\phi^{3}_{t}+\phi^{4}_{t}+t+\omega$}  \\
   6 & \multicolumn{1}{l}{$\phi_{t}+\phi^{2}_{t}+\phi^{3}_{t}+\phi^{4}_{t}+t+x$}  \\
   7 & \multicolumn{1}{l}{$\phi_{t}+\phi^{2}_{t}+\phi^{3}_{t}+\phi^{4}_{t}+t+y$}  \\
   8 & \multicolumn{1}{l}{$\phi_{t}+\phi^{2}_{t}+\phi^{3}_{t}+\phi^{4}_{t}+t+\omega+x$}  \\
   9 & \multicolumn{1}{l}{$\phi_{t}+\phi^{2}_{t}+\phi^{3}_{t}+\phi^{4}_{t}+t+\omega+y$}  \\
  10 & \multicolumn{1}{l}{$\phi_{t}+\phi^{2}_{t}+\phi^{3}_{t}+\phi^{4}_{t}+t+\omega+x+y$}  \\
  11 & \multicolumn{1}{l}{$\phi_{t}+\phi^{2}_{t}+\phi^{3}_{t}+\phi^{4}_{t}+t+\omega+\omega^2$}  \\
  12 & \multicolumn{1}{l}{$\phi_{t}+\phi^{2}_{t}+\phi^{3}_{t}+\phi^{4}_{t}+t+\omega$}  \\
  13 & \multicolumn{1}{l}{$\phi_{t}+\phi^{2}_{t}+\phi^{3}_{t}+\phi^{4}_{t}+t+x+x^2+y$}  \\
  14 & \multicolumn{1}{l}{$\phi_{t}+\phi^{2}_{t}+\phi^{3}_{t}+\phi^{4}_{t}+t+\omega+x+x^2+x^3$}  \\
  15 & \multicolumn{1}{l}{$\phi_{t}+\phi^{2}_{t}+\phi^{3}_{t}+\phi^{4}_{t}+t+\omega+y+y^2$       }  \\
  16 & \multicolumn{1}{l}{$\phi_{t}+\phi^{2}_{t}+\phi^{3}_{t}+\phi^{4}_{t}+t+\omega+y+y^2+y^3$       }  \\
  17 & \multicolumn{1}{l}{$\phi_{t}+\phi^{2}_{t}+\phi^{3}_{t}+\phi^{4}_{t}+t+\omega+\omega^2+x$       }  \\
  18 & \multicolumn{1}{l}{$\phi_{t}+\phi^{2}_{t}+\phi^{3}_{t}+\phi^{4}_{t}+t+x+x^2$       }  \\
  19 & \multicolumn{1}{l}{$\phi_{t}+\phi^{2}_{t}+\phi^{3}_{t}+\phi^{4}_{t}+t+x+x^2+x^3$}  \\
  20 & \multicolumn{1}{l}{$\phi_{t}+\phi^{2}_{t}+\phi^{3}_{t}+\phi^{4}_{t}+t+y+y^2$}  \\
  21 & \multicolumn{1}{l}{$\phi_{t}+\phi^{2}_{t}+\phi^{3}_{t}+\phi^{4}_{t}+t+y+y^2+y^3$}  \\
  22 & \multicolumn{1}{l}{$\phi_{t}+\phi^{2}_{t}+\phi^{3}_{t}+\phi^{4}_{t}+t+\omega+\omega^2+x$}  \\
  23 & \multicolumn{1}{l}{$\phi_{t}+\phi^{2}_{t}+\phi^{3}_{t}+\phi^{4}_{t}+t+\omega+\omega^2+x+y$}  \\
  24 & \multicolumn{1}{l}{$\phi_{t}+\phi^{2}_{t}+\phi^{3}_{t}+\phi^{4}_{t}+t+\omega+\omega^2+x+x^2+y+y^2$}  \\
  25 & \multicolumn{1}{l}{$\phi_{t}+\phi^{2}_{t}+\phi^{3}_{t}+\phi^{4}_{t}+t+\omega+\omega^2+x+x^3+y^2+y^3$}  \\
\enddata
\end{deluxetable}
\label{tab:sytematics}
\capstarttrue

\newpage
\capstartfalse
\startlongtable
\begin{deluxetable*}{cccccccc}
\tabletypesize{\scriptsize}
\tablewidth{0pt}
\tablecolumns{8}
\tablecaption{Systematics model selection for the STIS white light curves
\label{tab:systematic_models}}
\tablehead{\colhead{Model}  & \colhead{$\chi^{2}$} & \colhead{BIC}  & \colhead{$n$} & \colhead{d.o.f}  & \colhead{$i$ (degrees)} & \colhead{$a/R_{\star}$} & \colhead{$R_{p}/R_{\star}$}}
\startdata 
\sidehead{visit 52}
1  & 46.55 & 72.92  & 27 & 19 & 85.35 & 7.60 & 0.1677 \\
2  & 43.79 & 76.74  & 27 & 17 & 85.35 & 7.60 & 0.1676 \\
3  & 34.76 & 67.72  & 27 & 17 & 85.35 & 7.60 & 0.1688 \\
4  & 30.09 & 63.05  & 27 & 17 & 85.35 & 7.60 & 0.1691 \\
5  & 45.82 & 75.48  & 27 & 18 & 85.35 & 7.60 & 0.1676 \\
6  & 36.48 & 66.15  & 27 & 18 & 85.35 & 7.60 & 0.1685 \\
7  & 30.09 & 59.76  & 27 & 18 & 85.35 & 7.60 & 0.1691 \\
8  & 33.54 & 69.80  & 27 & 16 & 85.35 & 7.60 & 0.1683 \\
9  & 29.88 & 62.84  & 27 & 17 & 85.35 & 7.60 & 0.1690 \\
10 & 29.39 & 65.64  & 27 & 16 & 85.35 & 7.60 & 0.1690 \\
11 & 43.40 & 76.36  & 27 & 17 & 85.35 & 7.60 & 0.1675 \\
12 & 42.23 & 78.48  & 27 & 16 & 85.35 & 7.60 & 0.1673 \\
13 & 29.28 & 65.54  & 27 & 16 & 85.35 & 7.60 & 0.1693 \\
14 & 32.63 & 72.18  & 27 & 15 & 85.35 & 7.60 & 0.1687 \\
15 & 25.93 & 62.18  & 27 & 16 & 85.35 & 7.60 & 0.1696 \\
16 & 24.98 & 64.53  & 27 & 15 & 85.35 & 7.60 & 0.1696 \\
17 & 33.51 & 73.06  & 27 & 15 & 85.35 & 7.60 & 0.1683 \\
18 & 34.97 & 67.93  & 27 & 17 & 85.35 & 7.60 & 0.1689 \\
19 & 34.32 & 70.57  & 27 & 16 & 85.35 & 7.60 & 0.1689 \\
20 & 26.00 & 58.96  & 27 & 17 & 85.35 & 7.60 & 0.1696 \\
21 & 25.04 & 61.30  & 27 & 16 & 85.35 & 7.60 & 0.1695 \\
22 & 33.52 & 69.78  & 27 & 16 & 85.35 & 7.60 & 0.1683 \\
23 & 29.38 & 68.93  & 27 & 15 & 85.35 & 7.60 & 0.1690 \\
24 & 25.48 & 71.62  & 27 & 13 & 85.35 & 7.60 & 0.1696 \\
25 & 24.59 & 77.32  & 27 & 11 & 85.35 & 7.60 & 0.1697 \\
\sidehead{visit 53}
1 & 46.30 & 72.66 & 27 & 19 & 85.35 & 7.60 & 0.1656 \\
2 & 42.60 & 75.56 & 27 & 17 & 85.35 & 7.60 & 0.1655 \\
3 & 43.78 & 76.73 & 27 & 17 & 85.35 & 7.60 & 0.1654 \\
4 & 45.11 & 78.06 & 27 & 17 & 85.35 & 7.60 & 0.1658 \\
   5 & 43.80 & 73.46 & 27 & 18 & 85.35 & 7.60 & 0.1655 \\
   6 & 46.04 & 75.70 & 27 & 18 & 85.35 & 7.60 & 0.1657 \\
   7 & 46.09 & 75.75 & 27 & 18 & 85.35 & 7.60 & 0.1657 \\
   8 & 43.27 & 79.52 & 27 & 16 & 85.35 & 7.60 & 0.1656 \\
   9 & 43.70 & 76.66 & 27 & 17 & 85.35 & 7.60 & 0.1655 \\
  10 & 43.30 & 79.55 & 27 & 16 & 85.35 & 7.60 & 0.1656 \\
  11 & 43.61 & 76.57 & 27 & 17 & 85.35 & 7.60 & 0.1660 \\
  12 & 43.54 & 79.79 & 27 & 16 & 85.35 & 7.60 & 0.1659 \\
  13 & 44.95 & 81.21 & 27 & 16 & 85.35 & 7.60 & 0.1658 \\
  14 & 33.46 & 73.01 & 27 & 15 & 85.35 & 7.60 & 0.1659 \\
  15 & 41.64 & 77.90 & 27 & 16 & 85.35 & 7.60 & 0.1653 \\
  16 & 41.60 & 81.15 & 27 & 15 & 85.35 & 7.60 & 0.1653 \\
  17 & 43.25 & 82.80 & 27 & 15 & 85.35 & 7.60 & 0.1658 \\
  18 & 45.27 & 78.23 & 27 & 17 & 85.35 & 7.60 & 0.1657 \\
  19 & 40.70 & 76.95 & 27 & 16 & 85.35 & 7.60 & 0.1659 \\
  20 & 43.61 & 76.57 & 27 & 17 & 85.35 & 7.60 & 0.1654 \\
  21 & 43.33 & 79.59 & 27 & 16 & 85.35 & 7.60 & 0.1653 \\
  22 & 43.29 & 79.54 & 27 & 16 & 85.35 & 7.60 & 0.1658 \\
  23 & 43.26 & 82.81 & 27 & 15 & 85.35 & 7.60 & 0.1657 \\
  24 & 38.82 & 84.96 & 27 & 13 & 85.35 & 7.60 & 0.1658 \\
  25 & 26.00 & 78.73 & 27 & 11 & 85.35 & 7.60 & 0.1689 \\
\sidehead{visit 54}
   1 & 58.65 & 81.73 & 27 & 20 & 85.35 & 7.60 & 0.1681 \\
   2 & 50.54 & 80.20 & 27 & 18 & 85.35 & 7.60 & 0.1684 \\
   3 & 45.06 & 74.72 & 27 & 18 & 85.35 & 7.60 & 0.1671 \\
   4 & 54.79 & 84.45 & 27 & 18 & 85.35 & 7.60 & 0.1685 \\
   5 & 51.13 & 77.50 & 27 & 19 & 85.35 & 7.60 & 0.1682 \\
   6 & 48.52 & 74.89 & 27 & 19 & 85.35 & 7.60 & 0.1673 \\
   7 & 56.89 & 83.26 & 27 & 19 & 85.35 & 7.60 & 0.1681 \\
   8 & 38.76 & 68.42 & 27 & 18 & 85.35 & 7.60 & 0.1673 \\
   9 & 46.54 & 76.20 & 27 & 18 & 85.35 & 7.60 & 0.1681 \\
  10 & 35.87 & 68.83 & 27 & 17 & 85.35 & 7.60 & 0.1674 \\
  11 & 49.75 & 79.41 & 27 & 18 & 85.35 & 7.60 & 0.1681 \\
  12 & 51.13 & 77.50 & 27 & 19 & 85.35 & 7.60 & 0.1682 \\
  13 & 46.63 & 79.59 & 27 & 17 & 85.35 & 7.60 & 0.1678 \\
  14 & 33.56 & 69.82 & 27 & 16 & 85.35 & 7.60 & 0.1669 \\
  15 & 46.18 & 79.14 & 27 & 17 & 85.35 & 7.60 & 0.1681 \\
  16 & 45.65 & 81.90 & 27 & 16 & 85.35 & 7.60 & 0.1681 \\
  17 & 38.75 & 71.71 & 27 & 17 & 85.35 & 7.60 & 0.1673 \\
  18 & 47.33 & 76.99 & 27 & 18 & 85.35 & 7.60 & 0.1677 \\
  19 & 41.75 & 74.70 & 27 & 17 & 85.35 & 7.60 & 0.1672 \\
  20 & 56.75 & 86.41 & 27 & 18 & 85.35 & 7.60 & 0.1680 \\
  21 & 56.24 & 89.20 & 27 & 17 & 85.35 & 7.60 & 0.1681 \\
  22 & 38.75 & 71.71 & 27 & 17 & 85.35 & 7.60 & 0.1673 \\
  23 & 35.87 & 72.12 & 27 & 16 & 85.35 & 7.60 & 0.1674 \\
  24 & 30.89 & 73.73 & 27 & 14 & 85.35 & 7.60 & 0.1670 \\
  25 & 26.00 & 68.84 & 27 & 14 & 85.35 & 7.60 & 0.1675 \\
\enddata
\end{deluxetable*}
\label{tab:sytematic_models}
\capstarttrue

\newpage
\capstartfalse

\begin{deluxetable*}{ccccccc}
\tabletypesize{\scriptsize}
\tablewidth{0pt}
\tablecolumns{7}
\tablecaption{STIS G430L (visit 52) transmission spectrum results for WASP-52b\label{tab:v52}}

\tablehead{\colhead{$\lambda$ (\AA)}  & \colhead{$(R_{p}/R_{*})_{uncorr}$} & \colhead{$(R_{p}/R_{*})_{corr}$} & \colhead{$c_{1}$}  & \colhead{$c_{2}$}  & \colhead{$c_{3}$}  & \colhead{$c_{4}$} }

\startdata 
2900$-$3700   &   0.15814 $\pm$ 0.00712 & 0.17008 $\pm$ 0.00444 & 0.4371 & -0.7679 & 1.6319 & -0.3645 \\ 
3700$-$3950   &   0.16152 $\pm$ 0.00285 & 0.16801 $\pm$ 0.00294 & 0.7648 & -1.1573 & 1.7190 & -0.4157 \\
3950$-$4113   &   0.16997 $\pm$ 0.00300 & 0.16293 $\pm$ 0.00181 & 0.4778 & -0.6459 & 1.6564 & -0.5511 \\
4113$-$4250   &   0.16770 $\pm$ 0.00162 & 0.16633 $\pm$ 0.00128 & 0.4831 & -0.6750 & 1.7025 & -0.5879 \\
4250$-$4400   &   0.16683 $\pm$ 0.00151 & 0.16379 $\pm$ 0.00175 & 0.6151 & -0.8347 & 1.7798 & -0.6519 \\
4400$-$4500   &   0.16366 $\pm$ 0.00144 & 0.16739 $\pm$ 0.00175 & 0.4691 & -0.4858 & 1.5912 & -0.6570 \\
4500$-$4600   &   0.16610 $\pm$ 0.00151 & 0.16536 $\pm$ 0.00170 & 0.4777 & -0.4094 & 1.5029 & -0.6494 \\
4600$-$4700   &   0.16833 $\pm$ 0.00108 & 0.16574 $\pm$ 0.00158 & 0.5977 & -0.6932 & 1.6924 & -0.6811 \\
4700$-$4800   &   0.16453 $\pm$ 0.00191 & 0.16496 $\pm$ 0.00150 & 0.4411 & -0.2389 & 1.1548 & -0.4505 \\
4800$-$4900   &   0.16528 $\pm$ 0.00117 & 0.16532 $\pm$ 0.00134 & 0.5159 & -0.3136 & 1.2538 & -0.5560 \\
4900$-$5000   &   0.16317 $\pm$ 0.00159 & 0.16626 $\pm$ 0.00176 & 0.4399 & -0.1314 & 1.0138 & -0.4335 \\
5000$-$5100   &   0.16720 $\pm$ 0.00107 & 0.16696 $\pm$ 0.00135 & 0.5409 & -0.3989 & 1.1899 & -0.4566 \\
5100$-$5200   &   0.16898 $\pm$ 0.00178 & 0.16612 $\pm$ 0.00159 & 0.5605 & -0.4363 & 1.0910 & -0.3757 \\
5200$-$5300   &   0.16782 $\pm$ 0.00104 & 0.16525 $\pm$ 0.00103 & 0.5381 & -0.2913 & 1.0934 & -0.4736 \\
5300$-$5400   &   0.16556 $\pm$ 0.00116 & 0.16470 $\pm$ 0.00200 & 0.5732 & -0.3581 & 1.1751 & -0.5264 \\
5400$-$5500   &   0.16641 $\pm$ 0.00133 & 0.16719 $\pm$ 0.00122 & 0.5975 & -0.4018 & 1.1388 & -0.4758 \\
5500$-$5600   &   0.16496 $\pm$ 0.00110 & 0.16675 $\pm$ 0.00124 & 0.6370 & -0.4886 & 1.2206 & -0.5147 \\
5600$-$5700   &   0.16367 $\pm$ 0.00156 & 0.16527 $\pm$ 0.00101 & 0.5662 & -0.2476 & 0.9413 & -0.4076 \\
\enddata
\end{deluxetable*}
\label{tab:v52}
\capstarttrue

\newpage
\capstartfalse

\begin{deluxetable*}{ccccccc}
\tabletypesize{\scriptsize}
\tablewidth{0pt}
\tablecolumns{7}
\tablecaption{STIS G430L (visit 53) transmission spectrum results for WASP-52b\label{tab:v53}}
\tablehead{\colhead{$\lambda$ (\AA)}  & \colhead{$(R_{p}/R_{*})_{uncorr}$} & \colhead{$(R_{p}/R_{*})_{corr}$} & \colhead{$c_{1}$}  & \colhead{$c_{2}$}  & \colhead{$c_{3}$}  & \colhead{$c_{4}$} }
\startdata 
2900$-$3700   &  0.15814 $\pm$ 0.00712 & 0.15421 $\pm$ 0.00677 & 0.4371 & -0.7679 & 1.6319 & -0.3645 \\ 
3700$-$3950   &  0.16152 $\pm$ 0.00285 & 0.15781 $\pm$ 0.00296 & 0.7648 & -1.1573 & 1.7190 & -0.4157 \\
3950$-$4113   &  0.16997 $\pm$ 0.00300 & 0.16681 $\pm$ 0.00296 & 0.4778 & -0.6459 & 1.6564 & -0.5511 \\
4113$-$4250   &  0.16770 $\pm$ 0.00162 & 0.16457 $\pm$ 0.00159 & 0.4831 & -0.6750 & 1.7025 & -0.5879 \\
4250$-$4400   &  0.16683 $\pm$ 0.00151 & 0.16394 $\pm$ 0.00148 & 0.6151 & -0.8347 & 1.7798 & -0.6519 \\
4400$-$4500   &  0.16366 $\pm$ 0.00144 & 0.16101 $\pm$ 0.00142 & 0.4691 & -0.4858 & 1.5912 & -0.6570 \\
4500$-$4600   &  0.16610 $\pm$ 0.00151 & 0.16358 $\pm$ 0.00149 & 0.4777 & -0.4094 & 1.5029 & -0.6494 \\
4600$-$4700   &  0.16833 $\pm$ 0.00108 & 0.16583 $\pm$ 0.00106 & 0.5977 & -0.6932 & 1.6924 & -0.6811 \\
4700$-$4800   &  0.16453 $\pm$ 0.00191 & 0.16211 $\pm$ 0.00187 & 0.4411 & -0.2389 & 1.1548 & -0.4505 \\
4800$-$4900   &  0.16528 $\pm$ 0.00117 & 0.16290 $\pm$ 0.00115 & 0.5159 & -0.3136 & 1.2538 & -0.5560 \\
4900$-$5000   &  0.16317 $\pm$ 0.00159 & 0.16079 $\pm$ 0.00156 & 0.4399 & -0.1314 & 1.0138 & -0.4335 \\
5000$-$5100   &  0.16720 $\pm$ 0.00107 & 0.16463 $\pm$ 0.00105 & 0.5409 & -0.3989 & 1.1899 & -0.4566 \\
5100$-$5200   &  0.16898 $\pm$ 0.00178 & 0.16623 $\pm$ 0.00175 & 0.5605 & -0.4363 & 1.0910 & -0.3757 \\
5200$-$5300   &  0.16782 $\pm$ 0.00104 & 0.16541 $\pm$ 0.00103 & 0.5381 & -0.2913 & 1.0934 & -0.4736 \\
5300$-$5400   &  0.16556 $\pm$ 0.00116 & 0.16330 $\pm$ 0.00114 & 0.5732 & -0.3581 & 1.1751 & -0.5264 \\
5400$-$5500   &  0.16641 $\pm$ 0.00133 & 0.16416 $\pm$ 0.00132 & 0.5975 & -0.4018 & 1.1388 & -0.4758 \\
5500$-$5600   &  0.16496 $\pm$ 0.00110 & 0.16279 $\pm$ 0.00108 & 0.6370 & -0.4886 & 1.2206 & -0.5147 \\
5600$-$5700   &  0.16367 $\pm$ 0.00156 & 0.16157 $\pm$ 0.00154 & 0.5662 & -0.2476 & 0.9413 & -0.4076 \\
\enddata
\end{deluxetable*}
\label{tab:v53}
\capstarttrue

\end{document}